\documentclass[twocolumn]{aastex631}
\usepackage{amsmath}

\newcommand{\beq}{\begin{equation}}
\newcommand{\eeq}{\end{equation}}
\newcommand{\beqa}{\begin{eqnarray}}
\newcommand{\eeqa}{\end{eqnarray}}

\def\stacksymbols #1#2#3#4{\def\theguybelow{#2}
        \def\verticalposition{\lower#3pt}
        \def\spacingwithinsymbol{\baselineskip0pt\lineskip#4pt}
        \mathrel{\mathpalette\intermediary#1}}
\def\intermediary #1#2{\verticalposition\vbox{\spacingwithinsymbol
        \everycr={}\tabskip0pt
        \halign{$\mathsurround0pt#1\hfil##\hfil$\crcr#2\crcr
                \theguybelow\crcr}}}
\def\lta{\stacksymbols{<}{\sim}{2.5}{.2}}
\def\gta{\stacksymbols{>}{\sim}{2.5}{.2}}

\shorttitle{Multiphase gas in local massive elliptical galaxies}
\shortauthors{Temi et al.}
\begin{document}
\title{Probing multiphase gas in local massive elliptical galaxies via multiwavelength observations}

\author[0000-0002-8341-342X]{P. Temi}
\affiliation{Astrophysics Branch, NASA - Ames Research Center, MS 245-6, Moffett Field, CA 94035}

\author[0000-0003-2754-9258]{M. Gaspari}
\affiliation{INAF, Osservatorio di Astrofisica e Scienza dello Spazio, via Pietro Gobetti 93/3, 40129 Bologna, Italy }
\affiliation{Department of Astrophysical Sciences, Princeton University, 4 Ivy Lane, Princeton, NJ 08544-1001, USA}

\author[0000-0001-9807-8479]{F. Brighenti}
\affiliation{Dipartimento di Fisica e Astronomia, Universit\'a di Bologna, via Gobetti 93, 40127 Bologna, Italy}

\author[0000-0003-0392-0120]{N. Werner}
\affiliation{Department of Theoretical Physics and Astrophysics, Faculty of Science, Masaryk University, Kotl\'a\v{r}sk\'a 2, Brno, 611 37, Czech Republic}

\author[0000-0003-3471-7459]{R. Grossova}
\affiliation{Department of Theoretical Physics and Astrophysics, Faculty of Science, Masaryk University, Kotl\'a\v{r}sk\'a 2, Brno, 611 37, Czech Republic}

\author[0000-0002-0843-3009]{M. Gitti}
\affiliation{Dipartimento di Fisica e Astronomia, Universit\'a di Bologna, via Gobetti 93, 40127 Bologna, Italy}
\affiliation{INAF, Istituto di Radioastronomia, via Piero Gobetti 101, 40129 Bologna, Italy }

\author[0000-0001-5880-0703]{M. Sun}
\affiliation{Physics Department, University of Alabama in Huntsville, Huntsville, AL 35899, USA}

\author{A. Amblard}
\affiliation{Astrophysics Branch, NASA - Ames Research Center, MS 245-6, Moffett Field, CA 94035}
\affiliation{Carl Sagan Center, SETI Institute, 189 Bernardo Avenue, Mountain View, CA 94043, USA}

\author[0000-0002-9714-3862]{A. Simionescu}
\affiliation{SRON Netherlands Institute for Space Research, Niels Bohrweg 4, 2333 CA Leiden, The Netherlands }
\affiliation{Leiden Observatory, Leiden University, P.O. Box 9513, 2300 RA Leiden, The Netherlands}
\affiliation{Kavli Institute for the Physics and Mathematics of the Universe (WPI), The University of Tokyo, Kashiwa, Chiba 277-8583, Japan}

%\altaffiltext{\hspace{-0.15in} * }{E-mail: pasquale.temi@nasa.gov}

\begin{abstract}
\noindent
We investigate the cold and warm gas content, kinematics, and spatial distribution of six local massive elliptical galaxies to probe the origin of the multiphase gas in their atmospheres. We report new observations, including SOFIA [CII], ALMA CO, MUSE H$\alpha$+[NII]  and VLA radio observations. These are complemented by a large suite of multiwavelength archival datasets, including thermodynamical properties of the hot gas and radio jets, which are leveraged to investigate the role of AGN feeding/feedback in regulating the multiphase gas content. Our galaxy sample shows a significant diversity in cool gas content, spanning filamentary and rotating structures. In our non-central galaxies, the distribution of such gas is often concentrated, at variance with the more extended features observed in central galaxies. Misalignment between the multiphase gas and stars suggest that stellar mass loss is not the primary driver. A fraction of the cool gas might be acquired via galaxy interactions, but we do not find quantitative evidence of mergers in most of our systems. Instead, key evidence supports the origin via condensation out of the diffuse halo. Comparing with Chaotic Cold Accretion (CCA) simulations, we find that our cool gas-free galaxies are likely in the overheated phase of the self-regulated AGN cycle, while for our galaxies with cool gas the k-plot and AGN power correlation corroborate the phase of CCA feeding in which the condensation rain is triggering more vigorous AGN heating. The related C-ratio further shows that central/non-central galaxies are expected to generate an extended/inner rain, consistent with our sample. 
\vspace{+0.2cm}
\end{abstract}


\section{Introduction}

The cold/warm phase of the interstellar medium (ISM) in elliptical galaxies, although not as massive as in spirals, plays a key role in many, poorly understood physical processes which control the evolution of these objects.  The presence, distribution, and dynamics of the cold gas constrain the nature of the residual star formation in these galaxies, as well as the quenching mechanism and the black hole (BH) accretion$-$active galactic nucleus (AGN) feedback cycle.

Without an appropriate feedback mechanism (driven by AGN activity, star formation, or both) massive galaxies would grow to even larger stellar masses and intense star formation that would make these galaxies much bluer than they are \citep[e.g.][]{bower06, croton06}.
AGNs inject huge amounts of energy in the surrounding ISM in both radiative and mechanical form. 
Their impact on the host galaxies is demonstrated by the various type of outflows, now commonly detected in AGN hosts  \citep[e.g.][and references therein]{king15,mcnamara07}. 
Despite the strong negative feedback provided by the AGN, several studies of UV-optical colors indicate that a significant fraction ($\sim 30$\%) of the local early-type galaxies show evidence of star formation \citep{kaviraj07,sheen16}. The origin of the cold gas necessary to form stars is a matter of debate. While minor mergers could trigger some of the SF activity, internal sources of neutral/molecular material, like stellar mass loss or hot gas cooling, may well dominate the cold gas budget in ellipticals \citep{lagos14, valentini15, sheen16}.
To understand the critical role of hot gas cooling in the AGN feedback, information on the ISM cold gas phases becomes critical.

ALMA observations have detected CO emission and absorption in massive elliptical galaxies at the center of X-ray bright groups \citep{david14, temi18, rose19, maccagni21, north21}. For the archetypal group NGC 5044 the CO observations suggest that the molecular gas is indeed distributed in several clouds apparently orbiting within $\sim$2 kpc from the center. However, the [CII] image of the NGC 5044 core appears smoother and more extended than expected from the small number of discrete CO clouds, suggesting that some diffuse CO was resolved out by ALMA.
Colder molecular gas, possibly associated with [CII] emission from photodissociation regions, is also a common presence in central ellipticals \citep{werner14}. \\
H$\alpha$ emission is regularly detected in massive ellipticals \citep[e.g.][]{caon00, sarzi13, werner14} and probes warm $T \sim 10^4$ K gas, likely photoionized by UV radiation from post-AGB stars. The warm gas typically occupies a kinematically irregular region several kpc in size. 
One important feature observed in some, group centered, massive ellipticals \citep{werner14} is the correspondence between the cold molecular gas and the warm ionized phase. Such a co-spatial correlation of different phases can be understood in the multiphase cooling scenario \citep{valentini15, gaspari18}  if cooled clouds are characterized by inner layers of colder gas surrounded by an outer warmer skin made photo ionized by various local sources \citep[e.g.][]{mathews99, valentini15, gaspari17_rain}.

While a significant fraction of the cold gas mass in low to intermediate mass early-type (ET) galaxies is thought to have an external, merger-related origin \citep[e.g.][]{davis11,davis19},
in the most massive ET galaxies, of interest here, the cold gas phases are presumably generated internally by cooling  \citep{davis11,david14,werner14}.

In this paper we present new far-infrared [CII] observations of a sample of 6 local massive elliptical galaxies taken with the Field-Imaging Far-Infrared Line Spectrometer (FIFI-LS) on board of SOFIA \citep{young12, temi14}. 
The recently acquired SOFIA data are complemented with a large suite of multiwavelength observations available in literature which include, radio (ALMA, IRAM, VLA), optical ({\it HST}, SOAR and VMOS), IR ({\it Spitzer, Herschel}), and X-ray ({\it Chandra}) data. It is worth noting that, in addition to the new SOFIA observations,
some of the data retrieved from public archives have not been published yet, and forms a set of original and new observations. We will emphasize in the text any of these occurrences.
The paper is organized as follows. In section 2 we describe the sample, while the data reduction and analysis of the multiwavelength dataset are detailed in Section 3. The results for each individual galaxy are presented in section 4. The discussion and conclusions are presented in section 5 and 6, respectively.\\

\begin{table*}[ht!]
\footnotesize
\begin{center}
\caption{Galaxy sample information }
  \begin{tabular}{lcccccccccccc}
Galaxy  &      cz   &  D      &arcsec/pc &   Type    &   M$_K$ &  log(Re) & $\epsilon_e$ & $\lambda_{Re}$ & F/S &   $F_{\rm H\alpha}$                     &     ref H$\alpha$ & log(M$_H$) \\ 
                  & km/s   & Mpc    & &  &     mag    &  arcsec & & &  &  $\rm 10^{-13} erg ~cm^{-2} s^{-1}$      & & M$_\odot$\\ 
      (1)     &          (2)   &     (3)    & (4) &   (5)  &  (6) &   (7)     &  (8) & (9) & (10) & (11) & (12) & (13) \\
\hline
NGC~4203   &  1086  &  15.1  &  73 &    E-S0  (-2.7$\pm$0.7)  &  -23.44   &    1.47 & 0.154 &0.305 &F & 0.7     &       5  & 7.8$\pm$0.1   \\
NGC~4261   &   2212 &  31.6  & 152  &    E  (-4.8$\pm$0.4)  &  -25.18    &  1.58  &0.222&0.085&S &   4.1        &    1,5   & 7.9$\pm$0.1  \\   
NGC~4374 (M84)   & 1017  &  18.4  & 89  &    E  (-4.4$\pm$1.2)  &    -25.12  &     1.72 & 0.147& 0.024 & S &   3.8 &      2,5 & 8.2$\pm$0.3  \\
NGC~4406 (M86)  &   -224  &  17.1  & 83  &    E (-4.8$\pm$0.5)   &   -25.04  &    1.97  & 0.211&0.052&S &  11.4    &        2,3 & 4.7$\pm$0.4\\
NGC~4552 (M89)  &   344   &  15.8  & 76  &    E (-4.6$\pm$0.9)   &   -24.29  &   1.53   &   0.047 &0.049& S & 2.7  &    2,4,5,6 & 5.5$\pm$0.1  \\
NGC~4649 (M60)  &    1110  &  17.3  &  84 &    E (-4.6$\pm$0.9)    &    -25.46  &   1.82  &  0.156 & 0.127&F & 2.8  &        2   & 4.5$\pm$0.1  \\
NGC~4636$^a$  & 938 & 14.7 & 71 &  E (-4.8$\pm$0.5) & -24.36 & 1.95 & 0.036 & 0.25 & S & 2.7                        & 4 & 6.4$\pm$0.1 \\
 \end{tabular}
  \label{tab:gals}
  \end{center}  
 \tablecomments{ (1) : Galaxy name  ; (2) : velocity from NED ; (3) : distance from \cite{tonry01}; (4) : linear size conversion; (5) : Galaxy Type; Type T parameter in brackets from hyperleda\footnote{http://leda.univ-lyon1.fr/} (6) : absolute K band magnitude M$_K$ (2MASS keyword k\_m\_ext) from \cite{cappellari11}; (7) : effective radius R$_e$ from \cite{cappellari11}; (8,9,10) : ellipticity, specific stellar angular momentum, Fast or Slow rotator (Slow if  0.31$\sqrt{\epsilon_e}$/$\lambda_{Re} > 1$) at R$_e$; (11,12) : H$\alpha$+[NII] fluxes and references (1 :\cite{gavazzi18}; 2 : \cite{trinchieri91}; 3 : \cite{kenney08}; 4 : \cite{macchetto96}; 5 : \cite{ho97b}; 6 : \cite{boselli21}) (13) : gas mass estimated using a gas-to-dust ratio of 100 and the dust masses calculated by \cite{amblard14} using Spitzer and Herschel fluxes of these galaxies \citep{temi09,amblard14};$^a$:added to the sample to present new ALMA CO(3-2) data.
 }
\end{table*}

\section{Sample description}

The goal of the SOFIA FIFI-LS observations was to expand the sample of early type galaxies observed in the [CII] line with {\it Herschal} PACS by obtaining a complete coverage of the sample of 18 nearby ($d<100$~Mpc), massive, X-ray bright early-type galaxies, with declinations $\delta>0$, studied by \citet{dunn10}. These nearby massive galaxies are the lowest redshift proxies of the more distant cluster cooling cores. We focus on the nearest brightest systems in order to study AGN feedback in greatest possible detail. The galaxies span a broad range of X-ray morphologies and radio jet powers, making the sample well matched to exploring the parameter space of the radio mode AGN feedback cycle. 
The sample of \cite{dunn10} was derived from a larger catalogue of 530 ETGs selected by \cite{Beuing99} with morphological type $T \leq -2$ and considered 90\% complete at a magnitude of $B_T$ = 13.5 mag. The sample consists of all normal galaxies well representative of elliptical and lenticular galaxies, including galaxies at the centres of clusters and groups. 
From this initial catalogue of optically bright galaxies with X-ray luminosity measurements, all galaxies with 0.1--2.4 keV fluxes $ > 3 \times 10^{-12}\ {\rm erg \ s^{-1} \ cm^{-2}}$  and within a distance of 100 Mpc were down selected by \cite{dunn10}. This is essentially regarded as a complete sample of nearby elliptical galaxies with measured optical, X-ray and radio properties.
SOFIA FIFI-LS observations were obtained for the nearest 6 out of the 12 proposed galaxies.  
The six galaxies included here should be regarded as an extension of the earlier work initiated with Herschel PACS and expanded by SOFIA on the investigation of the origin of the cold gas phase in these systems. Since these galaxies have been selected from the original sample of \cite{dunn10},  the combined sample (Herschel + SOFIA) form a larger and more statistically significant sample. 
The complementarity of the two, as well as their differences -- Herschel's sample traces systems at the center of groups or clusters, whereas systems in the SOFIA's sample are not central -- is discussed throughout the paper.
These galaxies are also part of the ATLAS3D catalog \citep{cappellari11}, which
contains most of the massive ($L_{\rm K} > 8.2 \times 10^9
L_{\odot,{\rm K}}$, which translates to $\rm M_* \gta 6 \times
  10^9 M_\odot$) early-type galaxies in the northern hemisphere at a
distance lower than 42 Mpc.\\
Table \ref{tab:gals} shows general parameters for the sample under investigation.

\section{observations}
In this section we introduce the multiwavelength dataset available for the galaxy sample under investigation, starting with the new SOFIA Far-infrared [CII] observations. While data reduction and analysis is presented in detail for new and unpublished observations, we also briefly summarize the reduction processes and data analysis for archival and publicly available data and refer to the appropriate references in literature.

\subsection{[CII] SOFIA FIFI-LS Observations}
The sample of 6 galaxies was observed by 
FIFI-LS \citep{klein14,fischer18} on
board SOFIA for 10 to 30 minutes each between March,
12$^{\mathrm{th}}$ and October, 23$^{\mathrm{rd}}$ 2015 (see Table
\ref{tab:obs}) at an altitude between 12 and 13.5 km, depending
on the source.  In this paper, we use the red channel centered on the
velocity-corrected [CII] line (157.74 $\mu$m rest-frame) for each
galaxy, while the blue channel observed simultaneously the [OIII]
line (88.356 $\mu$m rest-frame).  The spectrometer is comprised of
25x16 pixels of Germanium Gallium-doped photoconductors.  An integral
field unit (IFU) rearranges the $5\times 5$ pixel FOV (each pixel is 12'' on
the sky for the red channel) into $25\times 1$ pixels that are dispersed by
a diffraction grating into the $25\times 16$ detector pixels. At the [CII]
wavelength the spectral resolution is about 250 km~s$^{-1}$ ($R\sim$1200) and
the spectral range is about 1500 km~s$^{-1}$. The observations were conducted
with 2-point symmetric chopping and with nodding in Nod Match Chop
(NMC) style with a standard ABBA nod cadence. The chopping frequency
was set to 2 Hz for NGC~4203, NGC~4261 and NGC~4649, and to 5 Hz
for NGC~4374, NGC~4406 and NGC~4552. NGC~4203 observations were slightly
dithered as well.  The data used in this paper have been processed by the
FIFI-LS pipeline, that subtracts chops, calibrates the wavelength and
position, applies a flat correction and combines grating scans. The
pipeline then produces Level 3 data by applying a telluric correction,
calibrating the fluxes, correcting the wavelength shifts and
resampling the wavelength.
At that level, the data is a
datacube of dimensions $25\times N_{\mathrm w}\times N_{\mathrm nod}$, where the first
dimension represents the $5\times 5$ pixel FOV, the second dimension the
spectral channel and the third dimension the number of nods performed
during the observations.  For NGC~4203, we used Level 4 data, where the
data has been spatially resampled and the nods combined since all the
nods do not point at the same location because of the dithering.
Level 4 data consists of a datacube of dimensions $N_{\mathrm x}\times N_{\mathrm y}\times N_{\mathrm w}$,
$N_{\mathrm x}$ and $N_{\mathrm y}$ are the number of re-sampled spatial pixels (2'' pixel
for the red channel) and $N_{\mathrm w}$ is the number of spectral channels.

\begin{table*}
\footnotesize
\begin{center}
 \caption{SOFIA FIFI-LS observation parameters of the galaxy sample.}
  \begin{tabular}{ccccccc}
    Galaxy & RA$_{tel}$ & DEC$_{tel}$ & Exposure (s) & Altitude (ft) & ZA (deg) & MISSION-ID \\
    \hline
NGC~4203 & 12h15m05.28s & +33d11m50.4s & 675.8 & 43002 & 64.6 &2015-10-23\_FI\_F250 \\
NGC~4261 & 12h19m23.16s & +05d49m31.0s & 1658.9 & 43504 & 36.1 &2015-03-13\_FI\_F200 \\
NGC~4374 & 12h25m03.72s & +12d53m12.8s & 952.3 & 40499 & 31.5 &2015-03-27\_FI\_F206 \\
NGC~4406 & 12h26m11.76s & +12d56m46.0s & 952.3 & 41015 & 33.5 &2015-03-27\_FI\_F206 \\
NGC~4552 & 12h35m39.84s & +12d33m23.0s & 1751.0 & 41018 & 37.9 &2015-03-27\_FI\_F206 \\
NGC~4649 & 12h43m40.08s & +11d33m10.1s & 890.9 & 42998 & 46.7 &2015-03-12\_FI\_F199 
  \end{tabular}
  \label{tab:obs}
  \end{center}
 \tablecomments{ The coordinates (RA,DEC) represent the position of the center of the
    telescope (center of the 5x5 pixel array as well), the altitude
    and zenith angle (ZA) are the mean values. The exposure represents
    the on-source time, total observation duration is general about
    2.5 time larger. The MISSION-ID indicates the observation date and
    the SOFIA flight number.
  }
\end{table*}

\subsubsection{[CII] SOFIA FIFI-LS data reduction}
Following recommendation from SOFIA staff, NGC~4203 data were analyzed
using Level 4 products, whereas the rest of the data was analyzed using
Level 3, because only NGC~4203 observations were dithered and benefit
from the regridding and interpolation that is taking place between
Level 3 and 4.\\
In order to detect emission lines in NGC~4203 data,
the spatial and spectral edges of the cube were first discarded, since
the noise level is significantly higher (less sampled) and these areas are more
prone to systematic effects. Emission lines are also not expected at
the edge of the spectrum nor at the edge of the observed area. In each
remaining spectral channel, pixels with a flux in excess of the median
flux by at least 4.45 times the median absolute deviation (equivalent
to 3-$\sigma$ but with outlier removal) are selected. The selected
pixels are clustered spatially and spectrally simultaneously using the
Density-based spatial clustering of applications with noise (DBSCAN)
algorithm implemented in the scikit-learn python
library\footnote{\url{https://scikit-learn.org}}, 3 pixels are required in
order to define a cluster and the maximum distance is set to 4,
otherwise all other parameters are set to default.\\ 
For each cluster, the brightest pixel is then identified (e.g. red cross in Figure 
\ref{fg:n4203_0} bottom plot) and used as the initial point to fit
the spatial distribution of the source with a 2D Gaussian function
(e.g. black contours in Figure \ref{fg:n4203_0} bottom plot). The fitted
Gaussian allows to measure the extent and position of the source and
allows to normalize the total line flux from the brightest pixel
spectrum.\\ 
The last step of the analysis of the SOFIA NGC~4203 data is to
fit the spectrum of the brightest pixel with a 1D Gaussian
function. We chose to fit the brightest pixel instead of integrating
over the source extent because of the poor signal-to-noise of the data.  The
spectra and their fits are presented in the top plots of Figure
\ref{fg:n4203_0}, along with the resulting parameters, and the amplitude
has been normalized by the source extent as indicated by the
units.\\

For the 5 other galaxies, Level 3 data were used. The data
were corrected for atmospheric transmission and all the observed
nods (approximately 30 to 60) of the 5x5 pixels were grouped together.
Each pixel is then processed in the following way. First a background
is fitted and removed for each of the observed spectra, using a Ridge
linear regression on channels away from the central channels (velocity
greater than 500 km/s) where we expect the emission lines but
excluding channels on the edges of the spectrum, since they are noisier. A
median spectrum is calculated from the observations of the pixels, that
fall in the central 90 percentile of a first estimate of the median
spectrum, and a standard deviation is calculated for each channel of the
spectrum.\\
A 1D Gaussian function is fitted on the median spectrum of
each of the 25 pixels. Fitted Gaussians with an absolute velocity
below 400 km/s and a signal-to-noise detection greater than 3 are
selected as emission line candidate.  The 2 candidates obtained through
this processing are presented in Figure \ref{fg:4261_cii} and \ref{fg:4406_cii}. Given the
sparsity of the spatial sampling (5x5 pixel image), the source is
assumed to be smaller than the FIFI-LS beam (the FIFI-LS beam size is used
for the flux density normalization).\\
For NGC~4261, given the width and the shape of the detected [CII]
feature, we performed a fit with 2 Gaussians (Figure \ref{fg:4261_cii}).
The resulting $\chi^2$ decreases significantly (by almost a factor 3),
and the Bayesian Information Criterion (BIC) is reduced by 13.9,
which strongly indicates that the two line model is favored. The velocity 
shift between the two lines is then 434 km/s and the velocity dispersion
is similar between the two lines.

\subsection{CO Observations}
We present new and archival CO observations taken with the 12 m array ALMA interferometer and with the single-dish IRAM 30m telescope.\\

\noindent
{\it ALMA Data} - Three galaxies in the sample, NGC~4261, NGC~4374, and NGC~4649  have been observed with ALMA.
In addition, we present new ALMA CO(3-2) observations of the central group elliptical galaxy NGC~4636.
Although this galaxy is not a member of the SOFIA [CII] galaxy sample, it has been included here because
of the newly acquired CO data.
NGC~4636 is part of the galaxy sample observed in the [CII] line by the {\it Herschel}
observatory \citep{werner14}.

We utilize ALMA CO(2-1) archival data for NGC~4261 acquired under the ALMA science program 2017.1.00301.S (PI A. Barth)(Figure \ref{fg:co21n4261}). The observations were taken with the 12m antenna interferometer in a configuration
providing 0.3$^{\prime \prime}$ angular resolution (maximum recoverable scale, MRSA, $~$ 4$^{\prime \prime}$)
and therefore probe smaller scales than [CII] data.
We also use ALMA CO observations of NGC~4374 and NGC~4649 presented in literature.
CO(2-1) emission in NGC~4374 is reported by \cite{boizelle17} using the 12 m array configured with 
a 0.3$^{\prime \prime}$ resolution.
A null detection is reported for NGC~4649 CO(3-2) from data recorded under the science program
2017.1.00830.S (PI N. Nagar)  that have an angular resolution of 0.12$^{\prime \prime}$. 

NGC~4636 has been observed with ALMA during Cycle 5 (project code:
2015.1.01107.S; PI: A. Simionescu). The ALMA interferometer was
configured such that its longest baseline was about 640 m and its
shortest baseline about 15 m. This configuration resulted in an
angular resolution of about 0.17'' and a maximum recoverable scale of
5.6''. Assuming a distance of 14.7 Mpc for NGC 4636, 0.17 and 5.6
arcseconds correspond to 12 and 399 pc respectively. All data were
taken in the ALMA band 7, one spectral window (spw) was centered
around the CO (3-2) line and three other spws measured the continuum.
The data were reduced using the CASA software \citep[version 4.7.2,
][]{mcmullin07}.  
The observations achieved a sensitivity of 0.71
mJy/beam in 10 km/s spectral windows for a 5322 second on-source time,
but no clear CO(3-2) line is detected in the data with a significance
greater than 5\,$\sigma$ using a similar analysis as in \cite{temi18}.
However a 3.5\,$\sigma$ point-like signal can be measured
at the location of a CO(2-1) cloud detected in \cite{temi18} near the
center of the galaxy. The inferred flux density CO(3-2) to CO(2-1) line ratio
is between 1.1 and 1.8 for this cloud. The source being unresolved in the CO(3-2) data
indicates that its size is smaller than $\sim$\,0.17'' (12 pc).\\

\noindent
{\it IRAM Data} - For NGC~4203, NGC~4406 and NGC~4552 CO(1-0) measurements were obtained with the IRAM 30m telescope, which has an
angular resolution of about 22$^{\prime \prime}$  and therefore probes angular
scales similar to SOFIA FIFI-LS for the [CII] line. \\

\subsection{Nuclear and Extended Dust }
We used archival data from the {\it Hubble Space Telescope} ({\it HST}) to
investigate the presence of dust in the core of each of the galaxy
sample.  In addition, cold dust emission in the extended
1$^\prime$--2$^\prime$ region has been probed in the infrared by the
{\it Spitzer Space Telescope} and the {\it Herschel Space Telescope}.

\subsubsection{Dust absorption from {\it HST}}

To estimate the amount and morphology of the dust absorption, we used
{\it HST} images collected on the Hubble Legacy Archive (HLA)
website \footnote{\url{https://hla.stsci.edu/}} in a blue/visible filter
(F450W, F547M, F555W depending on the galaxy) and red filter
(F814W). These data were observed as part of proposals \#5999,
\#11339, \#6094, \#5512, \#5454/6099, \#6286 with the Wide Field and
Planetary Camera 2 (WFPC2) and Advanced Camera for Surveys (ACS)
instruments.\\
The dust absorption on the ``blue'' filter data was
estimated by subtracting a model of the stellar emission. We attempted
to calculate the stellar emission model either with a functional form
obtained with
GALFIT\footnote{\url{https://users.obs.carnegiescience.edu/peng/work/galfit/galfit.html}}\citep{peng10}
or with the python library
photutils\footnote{\url{https://photutils.readthedocs.io/en/stable}}\citep{bradley19},
or with a linear function of the ``red'' filter assuming that the
galaxy SED does not change much across the galaxy. For GALFIT, we
first fitted a single component and increased the complexity up to
three components (not including a sky background component). We tried
several types of profile (Sersic, Nuker, Ferrer, etc.)  with GALFIT, but
our models from the photutils or the "red" filter seem to be closer to
the true stellar emission for these galaxies. In the end, we chose a
photutils model for NGC~4406 and NGC~4552, and a model estimated from
the red filter for the other galaxies.

 The dust absorption maps are presented in \S \ref{sec:results} for each individual galaxy in the sample 
 (Figure \ref{fg:n4203_1}, \ref{fg:4261_dust}, \ref{fg:4374_dust},  \ref{fg:4406_cii}, \ref{fg:4552_dust}, \ref{fg:4649_dust}). All galaxies
except NGC~4649 show some traces of dust. In these figures, the dust
absorption is seen in blue (darker blue indicates stronger
absorption), yellow and red indicate excess residual emission to the
stellar emission model, while green areas mark regions where the
stellar emission model is a good match to the galaxy flux (no
absorption, nor additional emission). The excess residual emission can
result from foreground sources or mismatch between the model and the
galaxy (as seen at the center of NGC 4406 and 4552). When using a red
filter as the stellar emission template, an excess residual emission
can come from sources that are bluer than the average stellar
population as well.\\

As shown in Figure \ref{fg:n4203_1} and Figure \ref{fg:4374_dust}, NGC~4203 and NGC~4374 have the
largest amount of dust with prominent dust lanes, visible in each
individual observation.  
 NGC~4261 has a very small (about 1$^{\prime \prime}$/150 pc diameter)
but dense disk of dust at its center, while NGC~4552 and NGC~4406 have
several dust plumes along the radial direction extending out up to
4$^{\prime \prime}$/300 pc away from the galaxy center.  The NGC~4406 dust image shows some
complex residual emission at its center, that could not be modelled
with the ellipsoid isophote model of photutils nor with a combination of
several sersic or exponential profiles with GALFIT. NGC~4649 seems to be
devoided of dust, no clear signal has been identified.
  
\subsubsection{FIR Emission from Dust}
{\it Spitzer} and {\it Herschel} provide data in the MIR and FIR with
the MIPS and SPIRE instruments \citep{rieke2004, griffin2010},
observing at 24, 70, 160, 250, 350 and 500 $\mu$m.
We used FIR dust maps and flux densities of the galaxy sample listed
in recent publications by \citet{temi2007a}, \citet{temi2007b} and
\citet{amblard14}.  The reader is referred to these publications to
obtain details on the data reduction and the generation of calibrated
images and derived flux densities.

\subsection{H$\alpha$+[NII] emission from {\it HST}}
 
We derived H$\alpha+\rm [NII]$ images for our sample galaxies, except for
NGC~4406 (no {\it HST} H$\alpha+\rm [NII]$ image) using {\it HST} archival data.  To
remove the continuum, a wide-band image centered off the
H$\alpha+\rm [NII]$ line was subtracted from the narrow-band image
centered on the H$\alpha+\rm [NII]$ line. The two filters were aligned,
cropped to an identical size and we used a linear regression to fit
out the continuum emission from the narrow-band filter (the regression
can take into account cross-calibration error and the steepness of the
galaxy SED, among other things).\\ The RANSAC Regressor algorithm from
the Python library scikit-learn\footnote{\url{https://scikit-learn.org}}
was chosen to perform the regression in order to reject outliers in
the data. The fit was also performed in a ring-shaped aperture to
remove the main contribution from the H$\alpha+\rm [NII]$ line
emission. The inner and outer radius of the ring was determined for
each galaxy by verifying that little to no H$\alpha+\rm [NII]$ emission
was present within the aperture (typical values for the inner and
outer radius are about 5 and 8 arcsecs).  The percentage of outliers
for each regression remains below 15\%, an acceptable level, and
mostly comes from point sources in the image that were not subtracted
away.\\ From the H$\alpha+\rm [NII]$ images, H$\alpha+\rm [NII]$ fluxes were
calculated by summing up the pixel flux in a circular aperture
centered on each galaxy. Assuming an uniform Gaussian noise, the
statistical error on the flux was calculated by multiplying the median
absolute deviation (scaled to the RMS) of the background pixels with
the square root of the number of pixels in the aperture.  In addition
we evaluated a systematic error, coming from our imperfect continuum
subtraction, by propagating the estimated error on the regression
coefficients (scaling factor with wide-band filter and background
value). Due to residual point sources affecting the H$\alpha+\rm [NII]$
images, it is likely that we have underestimated this error to some
degree.\\ H$\alpha+\rm [NII]$ emission, estimated from {\it HST} data, has been detected in
NGC~4203, NGC~4261 and
NGC~4374. The procedure described previously did not return any
significant signal for the two other galaxies for which data were
available (NGC~4552 and NGC~4649), we therefore decided not to include
the resulting images.  To compare the structure of H$\alpha+\rm [NII]$
with the structure of the dust in each galaxy, we overlaid a contour
plot of the dust of each galaxy on the H$\alpha+\rm [NII]$ image. All
images show extended H$\alpha+\rm [NII]$ emission and some level of
correlation with the dust absorption.

\subsection{H$\alpha$+[NII] emission from SOAR \& VMOS}
M84 was observed with MUSE on Feb. 6 and March 2, 2019, for four exposures of 600 sec each. NGC~4261 was observed with MUSE on April 17, 2016, for six exposures of 870 sec each. MUSE provides a spectroscopic data cube on a rectangular 1$'\times1'$ field, with a spectral coverage of 4800 - 9000 \AA. The MUSE data were reduced using the v2.8.1 ESO MUSE pipeline. Further sky subtraction was done with the Zurich Atmosphere Purge (ZAP) software \citep{soto16},
using two sky observations (600 sec each) taken along with the M84 exposures and empty sky positions for the NGC~4261 observations.
We then fit the emission lines ([NII]$\lambda$6548,6583, H$\alpha$$\lambda$6563, [SII]$\lambda$6716,6731)
using the KUBEVIZ code as in \citet{fossati16}. Groups of lines were fitted simultaneously using
a combination of 1D Gaussian functions with fixed relative velocities. The noise was measured from the
``stat'' data cube and renormalised on the line fluxes to take into account the correlated noise introduced
by resampling and smoothing, as extensively described in \citet{fossati16}.

\begin{table}[hb!]
\footnotesize
\caption{Summary of {\it Chandra}  observations.}
\centering
\begin{tabular}{lcccccccc}
\hline\hline
Galaxy  	                &	Obs. ID    & Obs. date	& Detector & Exp. (ks)    \\  
\hline 
NGC~4203                &  10535    &  2009-03-10  &  ACIS-S    &      42.1          \\
NGC~4261                &     834            &   2000-05-06     & ACIS-S    & 17.9        \\
                                &   9569            &   2008-02-12     & ACIS-S    &   87.6      \\
NGC~4374                &     803            &  2000-05-19     &   ACIS-S   &  25.9     \\     
NGC~4406                &      318          &    2000-04-07   &    ACIS-S  &     13.3    \\
NGC~4552                &      2072          &  2001-04-22      &  ACIS-S  &     44.0       \\
                                &       13985        &  2012-04-22        &  ACIS-S  &   41.2        \\
                                &     14358       &   2012-08-10    &   ACIS-S  &    41.5      \\
                                &     14359       &   2012-04-23    &   ACIS-S  &    44.0      \\
NGC~4636 		&	323		   &  2000-01-26      & ACIS-S	&	48.8 	\\
NGC~4649         &    785       &  2000-04-20      &  ACIS-S &   33.0    \\
                &    8182       &  2007-01-30      &  ACIS-S &   39.8    \\
                &    8507       &  2007-02-01      &  ACIS-S &   15.0    \\
                &    12975       &  2011-08-08      &  ACIS-S &   73.4    \\
                &    12976       &  2011-02-24      &  ACIS-S &   86.9    \\
                &    14328       &  2011-08-12      &  ACIS-S &   11.9    \\
\hline
\label{chandradata}
\end{tabular}
\end{table}

\subsection{X-ray emission, hot gas and entropy}

The reduction and analysis of the archival {\it Chandra} data follows the procedures described in \citet{werner2012} and \citet{lakhchaura2018}. The data were cleaned to remove periods of anomalously high background. The observations are summarised in Table~\ref{chandradata}. 

Background-subtracted images were created in six narrow energy bands, spanning 0.5--2.0~keV. These were flat fielded with respect to the median energy for each image and then co-added to produce the X-ray images shown in Fig.~\ref{fg:xrayflux}. 

In the first part of the spectral analysis, we measured the azimuthally averaged, deprojected radial distributions of densities, temperatures, and metallicities. We extracted spectra from concentric annuli with a signal to noise ratio of at least 18 ($\sim 320$ counts per region) and fitted them simultaneously in the 0.5--2~keV band using the PROJCT model implemented in XSPEC \citep{arnaud96}. The emission from each spherical shell was modelled with an absorbed APEC thermal model \citep{foster12} and all the deprojected densities ($n_{\rm e}$) and temperatures ($kT$) were determined simultaneously. We fitted two metallicity values for each galaxy, one for the shells outside the radius of $\sim 1$~kpc and another for the central regions of the galaxies. 

For galaxies with a sufficient number of X-ray photons, we produced 2D maps of projected thermodynamic quantities. The regions used for spectral mapping were determined using the contour binning algorithm \citep{sanders2006}, which groups neighbouring pixels of similar surface brightness until a desired signal-to-noise threshold is met. For the line-rich spectra of the relatively cool galaxies in our sample, we adopted a signal to noise ratio of 18 ($\sim 320$ counts per region), which allowed us to achieve better than 5 per cent accuracy in the temperature measurements. The spectral fitting was performed using the SPEX\footnote{\url{www.sron.nl/spex}} package \citep{kaastra1996}. The spectrum for each region was fitted in the 0.5--2.0~keV band with a model consisting of an absorbed single-temperature plasma in collisional ionization equilibrium, with the temperature and emission measure as free parameters. The absorption column densities, $N_{\rm H}$, were fixed to the values determined by the Leiden/Argentine/Bonn radio survey of HI \citep{kalberla2005} and the metallicities were fixed to 0.5 Solar \citep[see][for more details]{werner2012}. 

For all spectral fitting, we employ the extended C-statistics available in XSPEC and SPEX. All errors are quoted at the 68 per cent confidence level for one interesting parameter ($\Delta C = 1$).

\subsection{Radio emission}
The VLA radio observations in A, B, C and D configurations in L-band (at 1--2\,GHz centered at 1.5\,GHz) are reduced using {\sc NRAO} Common Astronomy Software Applications \citep[{\sc casa},][]{mcmullin07} version 4.7.2 and 5.6.1. 
Two categories of the data are analyzed depending on the year of their observation: either the historical VLA data which includes observations before an important VLA upgrade in 2011, or new Karl Jansky VLA data after this upgrade. The new observation data sets are available in the high-resolution VLA A configuration for sources NGC~4203 and NGC~4406 as well as in the more compact configurations C and D for NGC~4374, NGC~4552, NGC~4636, and NGC~4649. These new data were calibrated using the {\sc casa} pipeline version 1.3 and reduced by standard procedures as described in e.g. \cite{grossova2019}. On the other hand, the historical VLA observations were manually calibrated using the NRAO pre-upgrade calibration methods\footnote{\url{https://casaguides.nrao.edu/index.php/Jupiter:_continuum_polarization_calibration}}. The summary of observation details can be found in Table \ref{tab:radio_obs_details}.\\

For most targets, the model by \cite{perley2013} for standard VLA calibrator 3C~286 was used to determine the flux scales, except for NGC~4552 and NGC~4649, both in A configuration, which were calibrated with 3C~48. The final total intensity images, for both new Karl Jansky VLA and historical VLA observations (Figure \ref{fg:vla_nogas}, \ref{fg:vla_gas}), are created using the {\sc casa} MultiTerm MultiFrequency synthesis {\tt clean} algorithm \citep{rau11} with the {\tt briggs (robust=0)} weighting scheme \citep{briggs95}. When the dynamic range of the radio map (signal to noise ratio) reaches values of 100, the self-calibration method consisting of three cycles of phase and one cycle of amplitude and phase calibration, is performed. 
The VLA data reduction details and the total flux densities are presented in Table \ref{tab:radio_reduction_details}.

\begin{table}
\footnotesize
	\centering
\caption{The VLA observational details.}
	\label{tab:radio_obs_details}
%	\begin{threeparttable}
	\begin{tabular}{lccccr} % four columns, alignment for each
	\hline
		Galaxy & Config/Band & Obs. ID & Obs. date & TOS$^a$  \\
		\hline
NGC~4203 & A/L & 15A-305 & 2015-Jul-03 & 3600 \\
NGC~4261 & A/L & AL0693 & 2007-Jun-08/09 & 2190 \\
NGC~4261 & C/L & AL0693 & 2008-May-24/25 & 2190 \\
NGC~4374 & A/L & AB0920 & 1999-Jul-18 & 2970 \\
NGC~4374 & B/L & BW0003 & 1994-Aug-04 & 1180 \\
NGC~4374 & C/L & 14A-468 & 2014-Dec-24 & 5301 \\
NGC~4406 & A/L & 15A-305 & 2015-Jul-02 & 3600 \\
NGC~4552 & A/L & AC301 & 1991-Aug-24 & 240 \\
NGC~4552 & C/L & 16A-275 & 2016-Apr-04 & 2508 \\
NGC~4636 & A/L & AF0389 & 2002-Mar-12 & 12440 \\ 
NGC~4636 & C/L & 17A-073 & 2017-May-25 & 5319 \\
NGC~4649 & A/L & AC0301 & 1991-Aug-24 & 260 \\ 
NGC~4649$^b$ & B/L & AW0105 & 1984-Jan-24 & 4410 \\
NGC~4649 & D/L & 17A-073 & 2017-Jun-01 & 5295 \\

\hline\\
\end{tabular}
   \tablecomments{
      a) TOS: Time on source in seconds.  b) Previously published by \cite{dunn10}
   }
%   \end{threeparttable}
\end{table}

\begin{table*}
\footnotesize
\begin{center}
\caption{The VLA data reduction details of sample sources.}
\label{tab:radio_reduction_details}
	\begin{tabular}{lcccccc}
		\hline
	Galaxy & Config/Band & Restoring Beam & Position Angle & RMS noise & $S_{1.5\,{\rm GHz}}$ $\pm$ e$S_{1-2\,{\rm GHz}}$ & $P_{1-2\,{\rm GHz}}$ $\pm$ e$P_{1.5\,{\rm GHz}}$ \\
		
	 & & ($\arcsec\times\arcsec$) & (deg) & (Jy/beam)  & (Jy) & (W/Hz)\\
		\hline
NGC~4203 & A/L & 1.8$\times$0.9 & 72.5 & 1.1$\times 10^{-4}$ & $\left(7.8 \pm 0.3\right) \times 10^{-3}$ & $\left(21.8 \pm 0.9\right) \times 10^{21}$ \\
%NGC~4203 & D/L & 44.7$\times$41.0  & 2.8$\times 10^{-3}$ & - & - & -  \\
NGC~4261 & A/L & $1.4\times 1.3$ & -22.9 & 8.2$\times 10^{-6}$ & $\left(1.72 \pm 0.07\right) \times 10^{-2}$ & $\left(1.73 \pm 0.07\right) \times 10^{21}$ \\
NGC~4261 & C/L & $18.8\times 13.5$ & 51.8 & 1.4$\times 10^{-3}$ & $\left(1.27 \pm 0.05\right) \times 10^{-1}$ & $\left(1.28 \pm 0.05\right) \times 10^{22}$ \\
NGC~4374 & A/L & $1.5\times 1.3$ & 47.6 & $2.0\times 10^{-3}$ & $\left(4.78 \pm 0.23\right) \times 10^{-1}$ & $\left(1.96 \pm 0.09\right) \times 10^{22}$ \\
NGC~4374 & B/L & $4.6\times 4.4$ & -4.5 & 1.4$\times 10^{-3}$ & $\left(4.18 \pm 0.17\right) \times 10^{0}$ & $\left(1.71 \pm 0.07\right) \times 10^{23}$ \\ 
NGC~4374 & C/L & 38.8$\times$32.4 & -45.2 & 5.8$\times 10^{-3}$ & $\left(5.94 \pm 0.24\right) \times 10^{0}$ & $\left(2.43 \pm 0.10\right) \times 10^{23}$ \\
NGC~4406 & A/L & 1.1$\times$1.0 & -1.4 & 4.5$\times 10^{-5}$ & $\left(2.77 \pm 0.53\right) \times 10^{-4}$ & $\left(1.06 \pm 0.20\right) \times 10^{19}$ \\
%NGC~4406 & D/L &  45.0$\times$43.2 & 2.6$\times 10^{-4}$ & - & - & -   \\
NGC~4552 & A/L & $1.4\times 1.1$ & 13.6 & 2.7$\times 10^{-4}$ & $\left(5.97 \pm 0.25\right) \times 10^{-2}$ & $\left(1.83 \pm 0.08\right) \times 10^{21}$ \\
NGC~4552 & C/L & $11.4\times 10.4$ & -26.5 & 2.6$\times 10^{-4}$  & $\left(1.65 \pm 0.07\right) \times 10^{-1}$  & $\left(5.05 \pm 0.21\right) \times 10^{21}$ \\
NGC~4636 & A/L & $2.7\times 1.6$ & -74.7 & 4.4$\times 10^{-6}$ & $\left(5.95 \pm 0.24\right) \times 10^{-2}$ & $\left(1.81 \pm 0.07\right) \times 10^{21}$  \\
NGC~4636 & C/L & $13.8\times 10.7$ & 116.7  & 7.4$\times 10^{-5}$  & $\left(6.91 \pm 0.28\right) \times 10^{-2}$ & $\left(2.11 \pm 0.08\right) \times 10^{21}$ \\
NGC~4649 & A/L & $1.4\times 1.3$ & -8.5 & 2.2$\times 10^{-5}$ & $\left(1.44 \pm 0.06\right) \times 10^{-2}$ & $\left(4.69 \pm 0.19\right) \times 10^{20}$ \\
NGC~4649 & D/L & $11.5\times 9.1$ & -6.8  & 4.2$\times 10^{-5}$ & $\left(2.83 \pm 0.11\right) \times 10^{-2}$ & $\left(9.22 \pm 0.37\right) \times 10^{20}$\\
\hline
\end{tabular}
\end{center}
 \tablecomments{ Information derived from the radio total intensity maps produced with {\sc casa}. The table contains galaxy name, configuration and band, restoring beam size (resolution) and its position angle, RMS noise, total flux density ($S_{1-2\,{\rm GHz}}$) and radio power ($P_{1-2\,{\rm GHz}}$) at 1--2\,GHz.
 }
%	\label{tab:radio_reduction_details}
\end{table*}

\section{Results} \label{sec:results}
Table \ref{tab:CII_CO_gal} summarizes the results of the SOFIA [CII] and CO observations. Three measured fluxes in the 158 $\mu$m emission line are reported, along with three 3-$\sigma$ upper limits.  
The listed CO fluxes are 
obtained either from the literature or calculated in this paper using data in the ALMA science archive. Since CO(1-0) fluxes were not available for all the galaxies in the sample, we quoted the lowest energy CO transition available.\\
NGC~4261 CO(2-1) flux was calculated using data in the ALMA science archive and recently published by \cite{boizelle21}. Given the high angular resolution of the 12m array data, some diffuse CO component might have been filtered out by the interferometer such as was the case for NGC~5044 observations, where the ALMA CO(2-1) flux was only 20\% of the IRAM-30m flux \citep{david14}. The same caveat may apply for the CO(2-1) and CO(3-2) fluxes and upper limits of NGC~4374 and NGC~4649 as they were derived from ALMA data. The NGC~4374 flux was calculated by \cite{boizelle17} in the central 1$^{\prime \prime}$ radius of the galaxy (0.3$^{\prime \prime}$ resolution) although they noted that there is additional faint CO emission along the prominent dust lanes of this galaxy. The NGC~4649 CO(3-2) upper limit was calculated using the ALMA science archive. The CO upper limit fluxes are given for the ALMA beam size (generally sub-arcsec) and they correspond to 3-$\sigma$ in a 500 km/s window. References to the CO fluxes from literature are given in the table's caption.

\begin{table*}[ht!]
\footnotesize
\begin{center}
\caption{CII and CO emitting gas properties.}
\label{tab:CII_CO_gal}
  \begin{tabular}{lccccccccc}
    Galaxy  &   $F_{\rm CII (total)}$   &  $L_{\rm CII (total)}$ & v$_{CII}$ & FWHM$_{CII}$ &$F_{\rm CO} $ & v$_{CO}$ & FWHM$_{CO}$ & [CII]/CO(1-0) & $M_{\rm mol}$\\ 
NGC      &  $\rm 10^{-13} erg ~cm^{-2} s^{-1}$ & $\rm 10^{32} W$ & km/s & km/s &$\rm 10^{-17} erg ~cm^{-2} s^{-1}$ & km/s & km/s & $\rm 10^3$ & $\rm 10^{5} M_\odot$\\
\hline
4203   &   7.0  $\pm$  1.4     & 19 $\pm$ 4 &   -61, 82 & 188, 120 &3.7$\pm$0.4$^a$ & -90 & 150 & 18.9$\pm$4.3& 230$\pm$24\\
4261     &   2.5  $\pm$  0.3   & 30 $\pm$ 4 & 172 & 672 & 3.8$\pm$0.2$^b$ & -323,247& 342,318 & 18.4$\pm$2.4 &184$\pm$9\\
4261 (2G)  & 2.4 $\pm$ 0.3 & 30 $\pm$ 4 & -115, 319   & 310, 358 &  3.8$\pm$0.2$^b$ & -323, 247 & 342, 318 & 17.5$\pm$ 2.3 \\
4374      &   $<$ 0.5  &   $<$ 2.0 &  & &  3.7 $\pm$ 0.5$^c$& -- &-- & $<$ 4& 61$\pm$8\\
4406  &  0.9  $\pm$    0.2     & 3.1 $\pm$ 0.7 & -108 & 265 &  $<$ 3.5$^d$ &  & & $>$2.6& $<$279\\
4552     &   $<$ 0.6  & $<$ 1.8 &  &  &    $<$ 2.8$^e$ &  & & &$<$190 \\
4649    & $<$ 0.5       &   $<$ 1.8 & & & $<$ 0.2$^f$ &  & && $<$0.9\\
4636$^*$  &  1.05$\pm$0.04 & 2.7$\pm$0.1 & 22 & 361 &   0.25$\pm$0.01$^g$ & 210 & 61 &210$\pm$43 & 2.6$\pm$0.2
  \end{tabular}
  \end{center}
\tablecomments{ CII fluxes and luminosities obtained in this paper
  and in \cite{werner14} for NGC~4636$^*$. When no emission
  was detected, a 3-$\sigma$ upper limit flux per beam in a
  central 500 km/s window is quoted.
  CO fluxes obtained from the
    literature or ALMA science archive, the CO upper limit fluxes are
    given for ALMA beam size (generally sub-arcsec), upper limit corresponds
    to 3-$\sigma$ in a 500 km/s window. The [CII]/CO(1-0) ratio was calculated
    using \cite{vila03} observed line ratio.
    (a) :  CO(1-0) from \citet{welch03},
    (b) : CO(2-1) from this work using ALMA science archive, (c) : CO(2-1) from \citet{boizelle17},
    (d) : CO(1-0) from \citet{wiklind95}, (e) CO(1-0) from \citet{combes07},
    (f) : CO(3-2) from this work using ALMA science archive. (g) : CO(2-1) from \citet{temi18}
 }
%\label{tab:CII_CO_gal}
\end{table*}

\cite{vila03} found line ratios for elliptical galaxies on the scale of tens of arcseconds to be about 0.7$\pm$0.2 (2.8$\pm$0.9) for CO(2-1)/CO(1-0) in brightness temperature (flux) and 0.5$\pm$0.3 (4.5$\pm$3.0) for CO(3-2)/CO(1-0) in brightness temperature (flux). These CO line ratios were used to compute the [CII]/CO(1-0) ratio column in table \ref{tab:CII_CO_gal}; these ratios go from 2,600 (lower limit) to about 19,000.

Several observations in the Galaxy show a wide range of [CII]/CO(1-0) ratios depending on the type of regions observed, Photodissociation Regions (PDRs), HII regions, CO-dark H$_2$ cloud, etc. \citep[e.g.][]{wolfire89, pabst17,pineda13,langer14}.
On the theoretical ground, several factors have been scrutinized that could influence the amplitude of the [CII] emission line intensity with respect to other molecular lines. For instance, \cite{rollig06} examined the influence of the metallicity on the [CII] emission in PDRs and found that the [CII]/CO(1-0) ratio decreases very rapidly with increased metallicity. They also calculated the influence of the FUV incident radiation, the clump mass, the density and found that the [CII]-to-CO(2-1) ratio was increasing with FUV radiation, decreasing with clump mass and decreasing with density. Varying all these factors, they found that the [CII]/CO(1-0) ratio can vary from 10$^3$ to 10$^6$ in PDRs.\\
Given the large range of possible values, the many influencing factors and the uncertainty in our ratio measurements, it is hardly possible to draw strong conclusions from our galaxy samples. The observed ratios are within the expected range (observationally and theoretically); NGC~4203 and NGC~4261 are observationally on the larger side for PDRs, but could be explained by a dominant contribution from HII region or CO-dark H$_2$ cloud or by a low metallicity, high FUV background, small clump mass or density.

In the central regions of brightest group galaxies the line ratio H$\alpha+\rm [NII]/[CII]$ of these two co-spatial emissions usually spans a narrow range $\sim$ 1.5 - 2.5 with little radial variation for $r < 3$ kpc \citep{werner14}.
This evidence suggests that the low-excitation C$^+$ ion could exist in the warm gas, and
that a significant fraction of [CII] comes from the same gas that emits H$\alpha$, with the implication that the two gas phases have a common excitation mechanism \citep{werner14,canning16}.\\
Such a narrow range in the H$\alpha+\rm [NII]/[CII]$ ratio is not confirmed for the galaxies in our sample.
Among the galaxies with detected [CII] emitting gas, the ratio extends from $\sim 0.1$ in NGC~4203, to $\sim 12$ in NGC~4406, with NGC~4261 and NGC~4636 showing ratios of $\sim1.6$ and $\sim 2.5$, consistent with the findings of \cite{werner14}.
Moreover, the upper limits in the $157 \mu m$ line emission, reported for the remaining galaxies, imply H$\alpha+\rm [NII]/[CII]$ ratios
of $>7$ (NGC~4374), $>5.6$ (NGC~4649), and $>4.5$ (NGC~4552) that are inconsistent with values found in group-center galaxies.

\begin{table*}[ht!]
\footnotesize
\begin{center}
\caption{Galaxy sample FIR flux densities and luminosities.}
  \label{tab:dustgals}
  \begin{tabular}{cccccccc}
    Galaxy  &   $F_{70\mu m}$  &  $F_{160\mu m}$& $F_{250\mu m}$  & $F_{350\mu m}$  & $F_{500\mu m}$  & $Log(L_{\rm d})$ &$ Log(M_{\rm d})$   \\ 
            &  (mJy) & (mJy) & (mJy) & (mJy) & (mJy) &$L_\odot $ & $M_\odot $ \\
\hline
NGC4203   &   933  $\pm$111  & 2701$\pm$259  & 1153$\pm$173& 526$\pm$84& 191$\pm$30 & 8.55$\pm$0.10 & 5.79$\pm$0.14    \\
NGC4261     &   127  $\pm$ 11  & 375$\pm$18 & 237$\pm$36& 309$\pm$57& 216$\pm$33 & 8.52$\pm$0.02  & 5.94$\pm$0.05   \\
NGC4374      &   67$\pm$9 & 535$\pm$61 & 239$\pm$36& 146$\pm$22& 119$\pm$18 & 8.46$\pm$0.11  &6.21$\pm$0.26  \\
NGC4406  &  64  $\pm$8  & 90$\pm$12    & 1$\pm$27 & 7$\pm$27& 8$\pm$20 & 5.04$\pm$0.03 &2.65$\pm$0.42  \\
NGC4552     &  96$\pm$10  & 188$\pm$16  & 2$\pm$27& 21$\pm$4& 18$\pm$16 & 6.63$\pm$0.03 &3.50$\pm$0.06    \\
NGC4649    & 48$\pm$7    & 0$\pm$30  & 1$\pm$31& 0$\pm$24& 0$\pm$19 & 8.32$\pm$0.03 & 2.52$\pm$0.03 \\
NGC4636$^*$  &  197$\pm$12 & 185$\pm$24  & 88$\pm$13 & 33$\pm$54 & 12$\pm$15 & 7.74$\pm$0.02 &4.36$\pm$0.07  
  \end{tabular}
  \end{center}
\tablecomments{FIR flux densities and luminosities are from {\it Spitzer} and {\it Herschel}. Dust masses from SED modeling  are from \citet{amblard14}.
 }
%  \label{tab:dustgals}
\end{table*}

Table \ref{tab:dustgals} presents the integrated flux density at all the FIR bands along with the dust luminosity $L_{\rm d}$ and mass $M_{\rm d}$.  Dust mass and luminosity have been evaluated using the spectral energy distribution (SED) modeling software CIGALEMC \citep{serra2011} and MAGPHYS14 \citep{dacunha2008}, and by fitting some modified black-body spectra to the FIR portion of our data ($\lambda > 60\mu m$) (see \citet{amblard14} for details).  Both SED fitters are capable of modeling the galaxy SED from a UV to a millimeter wavelength and constraining the computed $L_{\rm d}$ and $M_{\rm d}$ by fitting the absorption in UV, optical and the FIR emission consistently.  Given the excellent sampling of the FIR SED for all the galaxy sample, the computed dust luminosity and mass are well constrained.  It is worth noting that, although all the galaxies have similar $L_B$, the dust luminosity ranges over more than 2 orders of magnitude. This is also reflected in the total dust content: on one extreme NGC~4649 is consistent with a dust--free galaxy, while NGC~4261 accounts for $\sim 10^6 M_\odot$ of dust.  

Below, we report the results from the new SOFIA and ALMA observations, combined with archival multiwavelength data, on an object-by-object basis. 
The outlined  physical conditions of the ISM in each object will provide a baseline for the discussion and interpretation of the data.
 It should be noted that the units of $\Delta$RA and $\Delta$Dec in all the maps throughout the manuscript are expressed in units of arc-seconds.

\begin{figure}
  \centering
   \includegraphics[width=8.5cm]{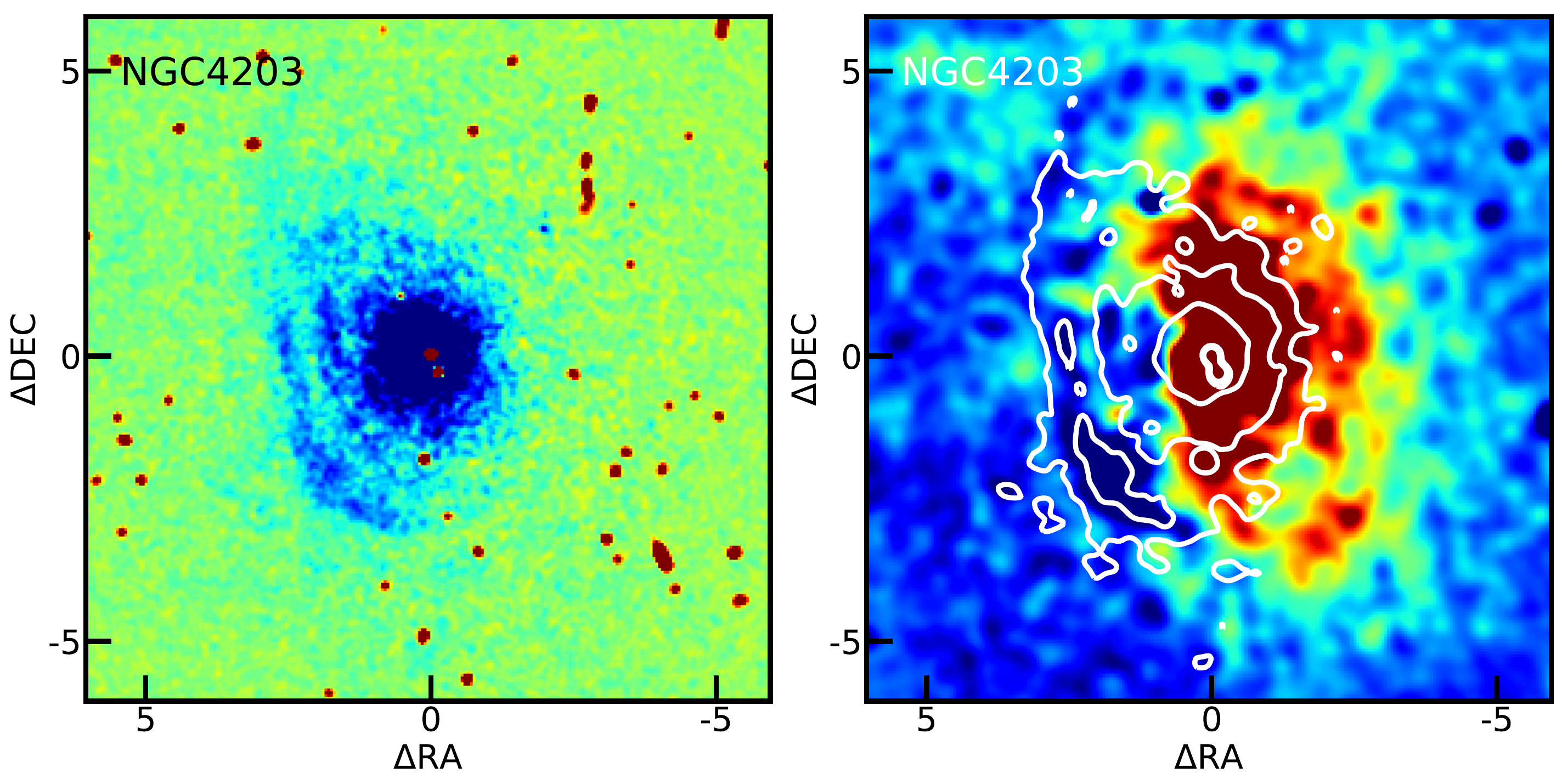}
   \caption{{\bf Left :} NGC~4203 dust absorption map from {\it HST}: a
    stellar emission model obtained from either a ``red'' filter (F814W) or a functional form (ellipsoidal isophote from photutils or radial profile from GALFIT) is subtracted to the ``blue'' filter F555W observation. The dust absorption appears in blue in this color scale, zero emission is green, and red area represent residual emission or emission bluer than the average stellar population in the galaxy. {\bf Right :} {\it HST} $H\alpha + [NII]$ map: a stellar emission continuum is removed from a narrow filter. The dust absorption is overlaid with white contours.}
\label{fg:n4203_1}
\vspace{+0.2cm}
\end{figure}

\subsection{NGC~4203}

The early-type galaxy NGC~4203 has been classified as a fast rotator by \citet{emsellem11} and consistently shows a central disk of dusty gas.  
The {\it HST} dust map of the central $\sim 10^{\prime \prime}$ shows a dusty central core
with one or two arms in the South-North direction compatible with its
face-on orientation.  The central disk of dust extends $\sim$350 pc in diameter (Figure \ref{fg:n4203_1}) and the filamentary structure is reminiscent of possible spiral arms.

\begin{figure}[ht!]
  \centering
   \includegraphics[width=8.5cm]{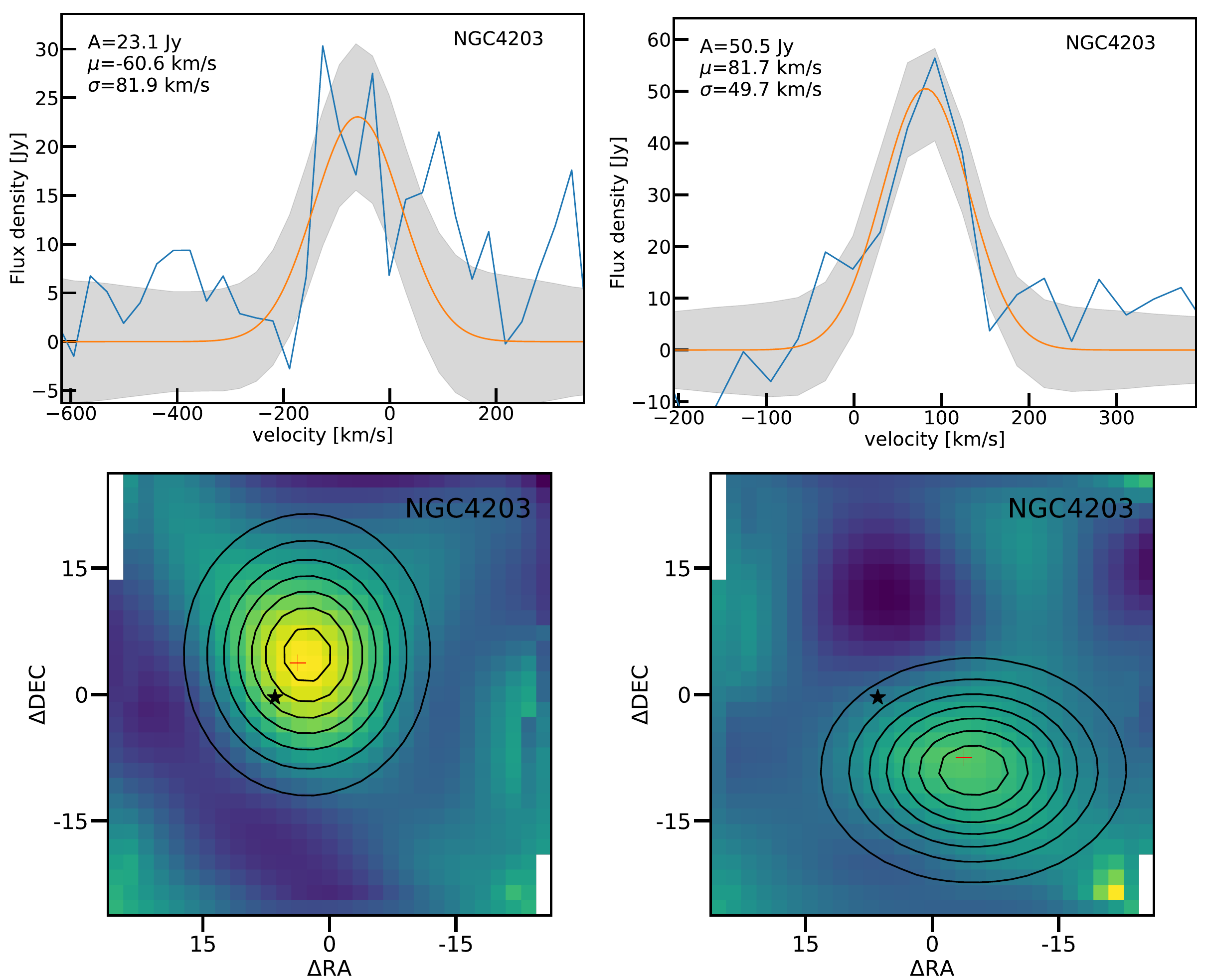}
  \caption{{\bf Top :} NGC~4203 CII lines (blue solid lines) detected in SOFIA FIFI-LS. The orange solid lines
    represent a Gaussian fit to the line, the parameters of the Gaussian are indicated in top left corner (A: amplitude, $\mu$ :
    velocity center, $\sigma$ : velocity width), and the name of the galaxy in the top right corner. The grey shade areas
    indicate the noise level for each spectrum. {\bf Bottom :} Images at the wavelength of the line maximum emission for the
    two observed lines of NGC~4203, the black star indicate the position of the galaxy (NED coordinates), the black contours
  represent a 2D-Gaussian fit to the line source.}
\label{fg:n4203_0}
\end{figure}

The galaxy has a strong integrated far-infrared flux with a peak at 160 $\mu$m of $\sim$ 2.7 Jy \citep{amblard14}.
Deep H~{\small I}  observations reveal a large gas reservoir ($ \sim 10^9 M_\odot$),  distributed in a disc with two distinct components  \citep{yildiz15}. 
Both [CII] emission and CO(2-1) emitting gas have been detected at a systemic velocity of $\sim$ -70 km/s and velocity dispersion of $\sim$150 km/s. The new SOFIA data reveal two distinct [CII] emission lines (Figure \ref{fg:n4203_0}). The detections are compact and unresolved by the FIFI-LS instrument and localized within the central 2 kpc.
The central H$\alpha$+[NII] gas emission, as revealed by the {\it HST} map (Figure \ref{fg:n4203_1}), shows the ionized gas
distributed in an ellipsoid of about $7^{\prime \prime}$ major axis along the
North-South direction with a larger emission in its western section.
The lack of H$\alpha$+[NII] emission in the Eastern side of the galaxy is
correlated with arm-like features in the dust absorption estimate presented in Figure \ref{fg:n4203_1}.

The X-ray observations of NGC~4203 are not very deep and suffer from a low number of detected counts when compared to the other galaxies (Figure \ref{fg:xrayflux}). 
The X-ray morphology appears relatively spherically symmetric.
However the data, in combination with older observations using the {\it Einstein X-ray Observatory} \citep{fabbiano1992} show evidence of ram pressure stripping with a tail extending to the South. 
The  nuclear radio emission observed at 1--2\,GHz taken in several VLA configurations shows a 
weak point source radio emission (Figure \ref{fg:vla_gas}; top left). 

\subsection{NGC~4261}
NGC~4261 is a giant elliptical E2 (from RC3) galaxy also known as the
radio galaxy 3C270 \citep{birkinshaw85} and is the largest member of a
group of 33 galaxies \citep{nolthenius93}.

\begin{figure}[ht!]
  \centering
  \includegraphics[width=8.5cm]{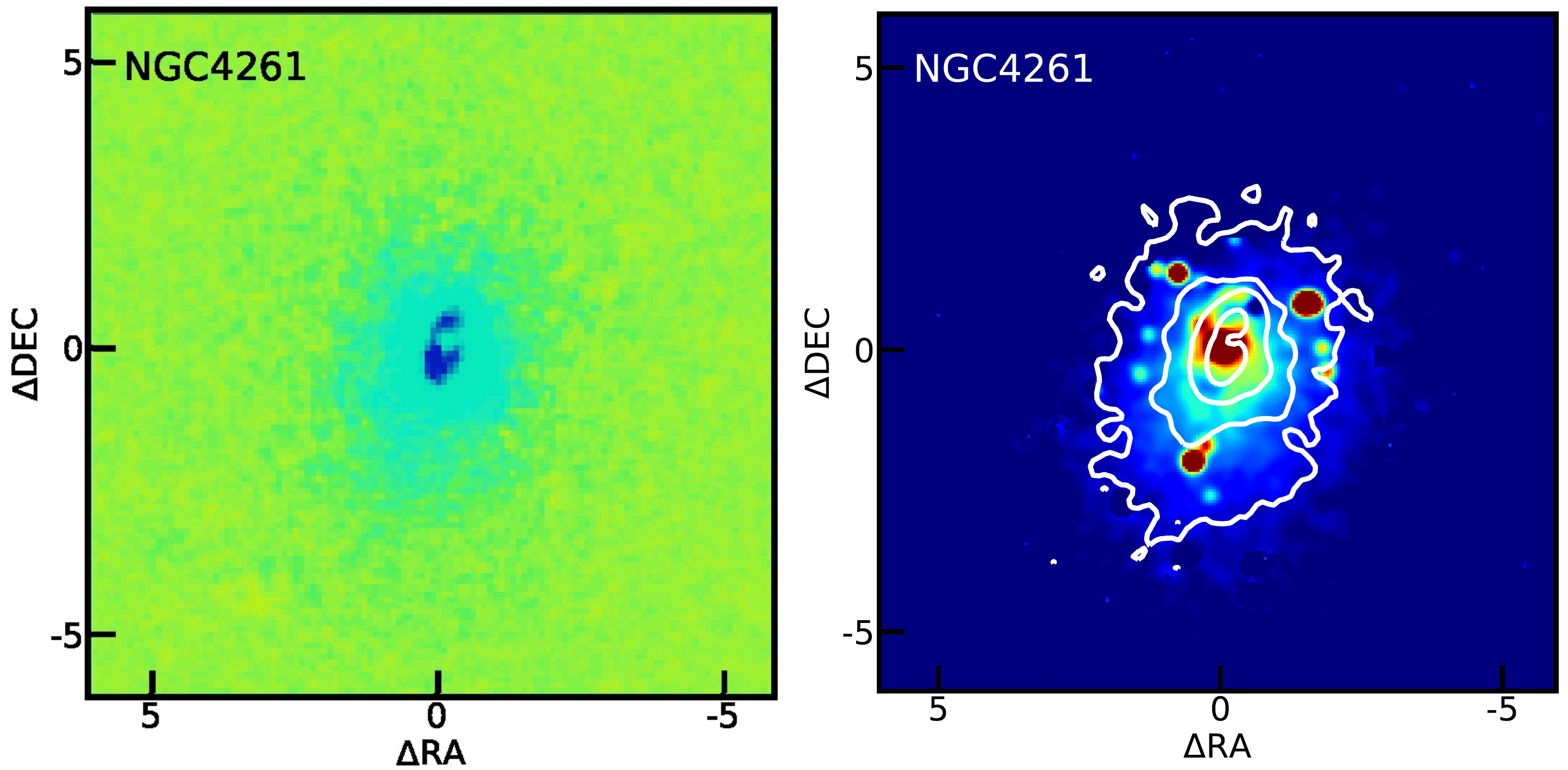}
  \caption{{\bf Left:} {\it HST} dust absorption map of NGC~4261.  The central disk of dust
    appears in dark blue in this color scale, zero emission is green. {\bf Right:} {\it HST} $H\alpha + [NII]$ map: a stellar emission continuum is removed from a narrow filter. The dust absorption is overlaid with white contours. }
\label{fg:4261_dust}
\end{figure}

\begin{figure}[ht!]
  \centering
  \includegraphics[width=8.5cm]{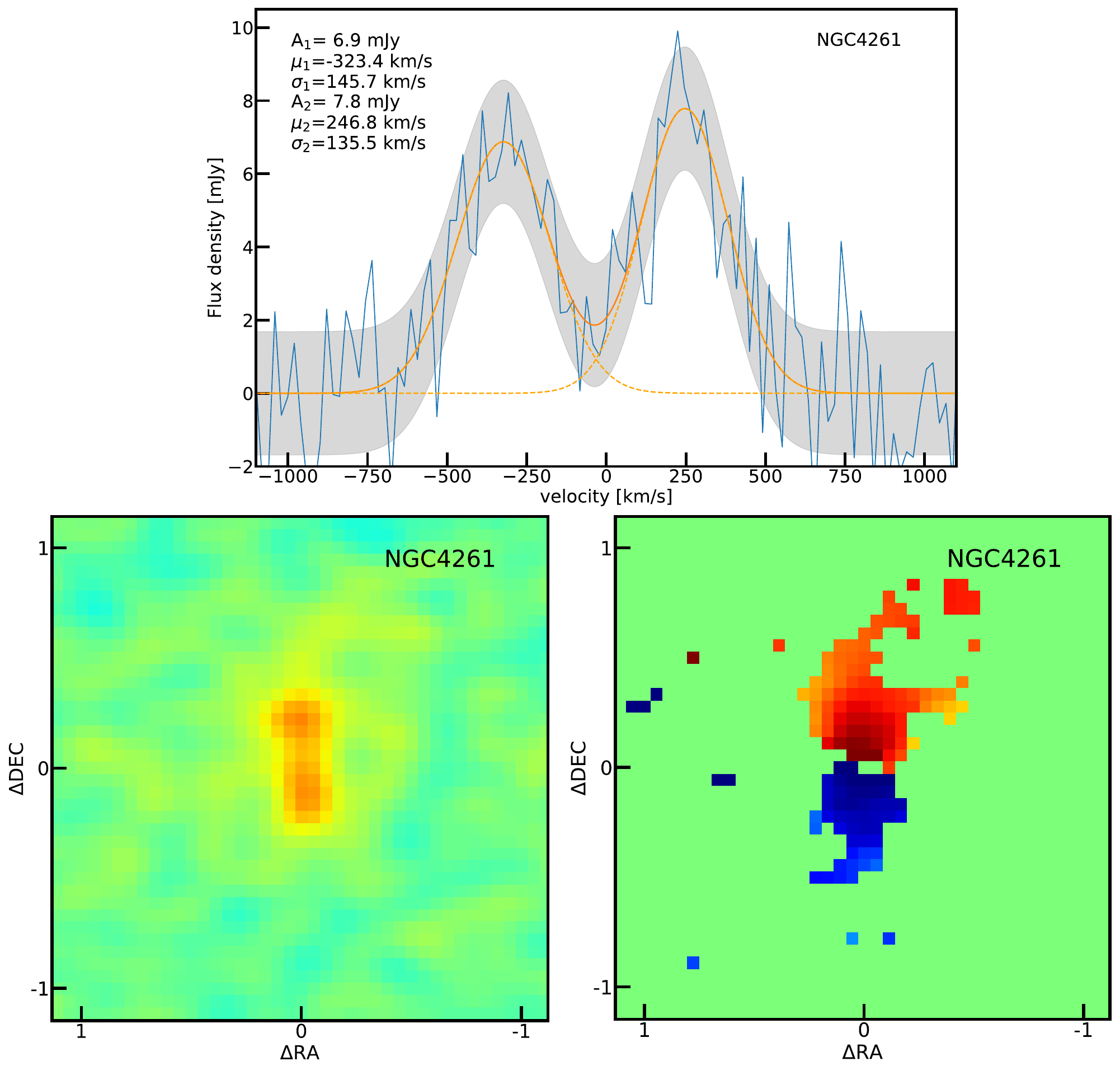}
  \caption{{\bf Top :} spectrum of CO(2-1) emission lines (blue solid line) observed by ALMA (about 0.3" angular resolution) with a 1" radius aperture. The orange solid lines
    represent a Gaussian fit to the lines. Parameters of the Gaussians are indicated in the top left corner (A: amplitude, $\mu$ :
    velocity center, $\sigma$ : velocity width), and the name of the galaxy in the top right corner. The grey shade area indicates the noise level. {\bf Lower left:} intensity of the CO(2-1) emission integrated over the velocity. {\bf Lower right:} velocity of the CO(2-1) emission line, the color scale goes from -350 to 350 km/s (blue to red).
    }
\label{fg:co21n4261}
\end{figure}

\begin{figure*}[ht!]
  \centering
  \includegraphics[width=12.5cm]{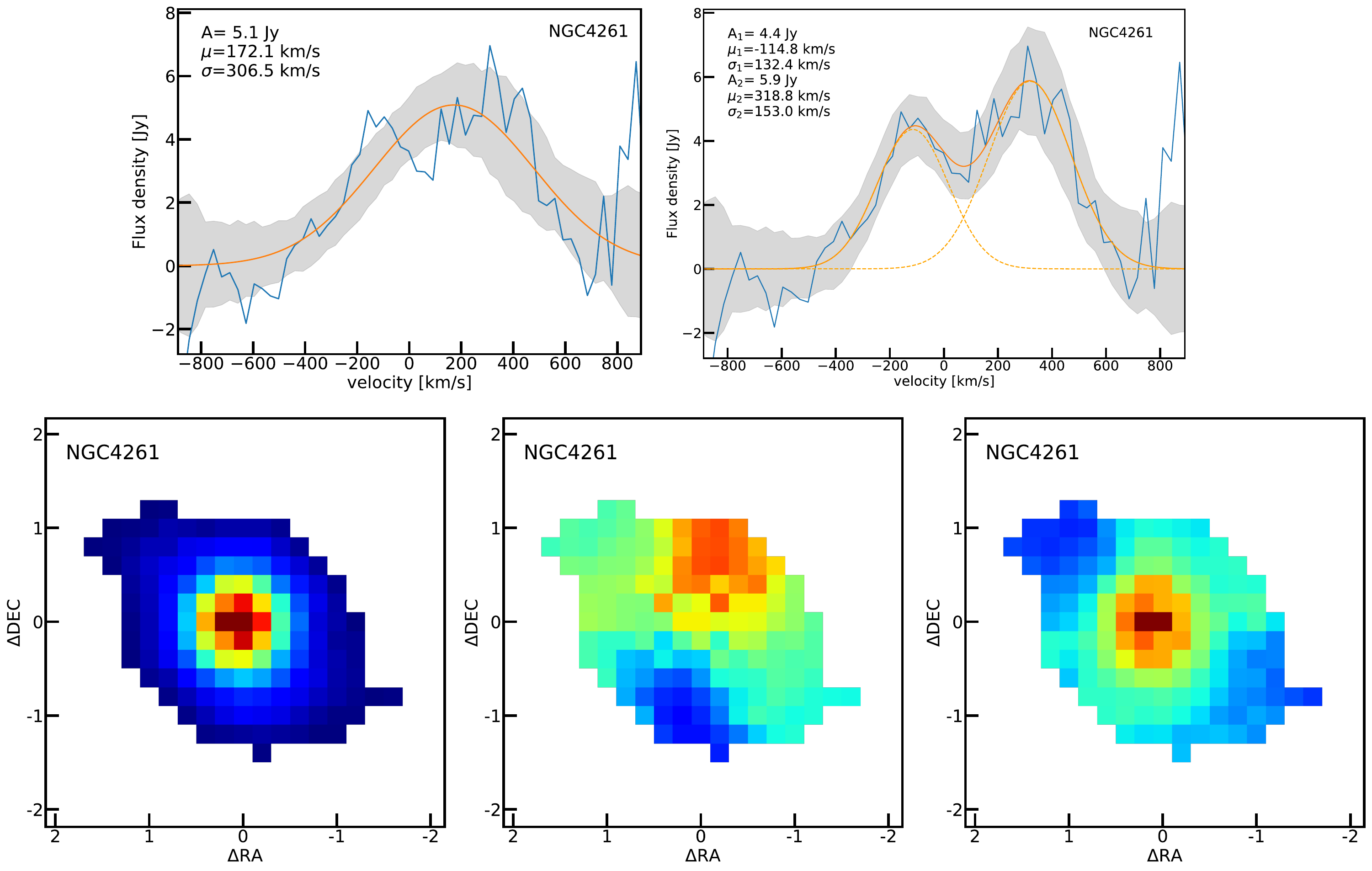}
  \caption{{\bf Top:} CII lines (blue solid lines) detected with SOFIA FIFI-LS in NGC~4261. The orange solid lines
    represent a Gaussian fit to the line, the parameters of the Gaussian are indicated in top left corner (A: amplitude, $\mu$ :
    velocity center, $\sigma$ : velocity width), and the name of the galaxy in the top right corner. The grey shade areas
    indicate the noise level for each spectrum. The top right panel shows an alternative fit for NGC~4261 with 2 Gaussian lines instead of one.
    {\bf Bottom:}
    H$\alpha$ emission (left), velocity (center) and velocity
    dispersion (right) estimated on MUSE spectra for NGC~4261. On the color scale, the velocity ranges
    from -150 to 150 km/s, the velocity dispersion range from 0 to 350
    km/s. The H$\alpha$ image is in units of 10$^{-20}$ erg cm$^{-2}$ s$^{-1}$. The rotation amplitude (300 km/s)
    of NGC~4261 ionized gas visible in the H$\alpha$ seems to be of
    the same order as the peak difference (400 km/s) in the [CII]
    observations, the orientation can not be probed with [CII] due to
    the small extent of the gas distribution. The velocity dispersion
    is also of the same order (about 200 km/s FWHM). These
    similarities could hint a common origin for these two lines.}
\label{fg:4261_cii}
\end{figure*}

The {\it HST} dust absorption map of NGC~4261 (Figure \ref{fg:4261_dust}) clearly shows a central disk of dust that extends $\sim$250 pc. Its size and location are well matched by the CO(2-1) data from ALMA (Figure \ref{fg:co21n4261}). The interferometric data  indicate a molecular gas rotation in the south to north direction of the molecular gas with a peak to peak amplitude of -350 to 250 km/s that is similar to the dynamics of the ionized gas measured by the H$\alpha$ emission. The peak to peak rotation amplitude is also similar to the [CII] measurement albeit a bit larger (amplitude of about 430 km/s for [CII] versus 570 km/s CO(2-1)) (Figure \ref{fg:4261_cii}).

The {\it HST} H$\alpha$+[NII] emission is distributed in a very compact, about $1^{\prime \prime}$ major
axis ring and a strong point-like central emission. The ring is
elongated in the North-South direction similarly to the dust disk, visible in Figure \ref{fg:4261_dust}. Both have a similar spatial extent, the dust absorption seems to be a bit more extended, but it could be due to a difference in the signal-to-noise ratio between the dust absorption and the H$\alpha$+[NII] emission estimates.

From the MUSE spectra (Figure \ref{fg:4261_cii}) the H$\alpha$ emitting gas exhibits a velocity range from -150 to 150 km/s. The rotation amplitude (300 km/s) of the ionized gas seems to be of the same order as the peak difference ($\sim$ 430 km/s) in the [CII] observations, the orientation cannot be probed with [CII] due to the small extent of the gas distribution. The velocity dispersion is also of the same order (about 200 km/s FWHM). These similarities could hint a common origin for these two lines.\\

The X-ray morphology appears relatively spherically symmetric (Figure \ref{fg:xrayflux}).
The radio emission observed at 1--2\,GHz reveals powerful jets penetrating the atmosphere and depositing their energy at large radii (Figure \ref{fg:vla_gas}; top middle).
The disk of dust, as revealed by {\it HST} data, is aligned with its radio axis and perpendicular to the stellar
component of the galaxy \citep{jaffe93}.

\subsection{NGC~4374}
NGC~4374 (M84, also known as the FRI radio
galaxy 3C272.1) is a giant elliptical E1 (from RC3) galaxy located in
the core region of the Virgo cluster. 

\begin{figure}[ht!]
  \centering
  \includegraphics[width=8.5cm]{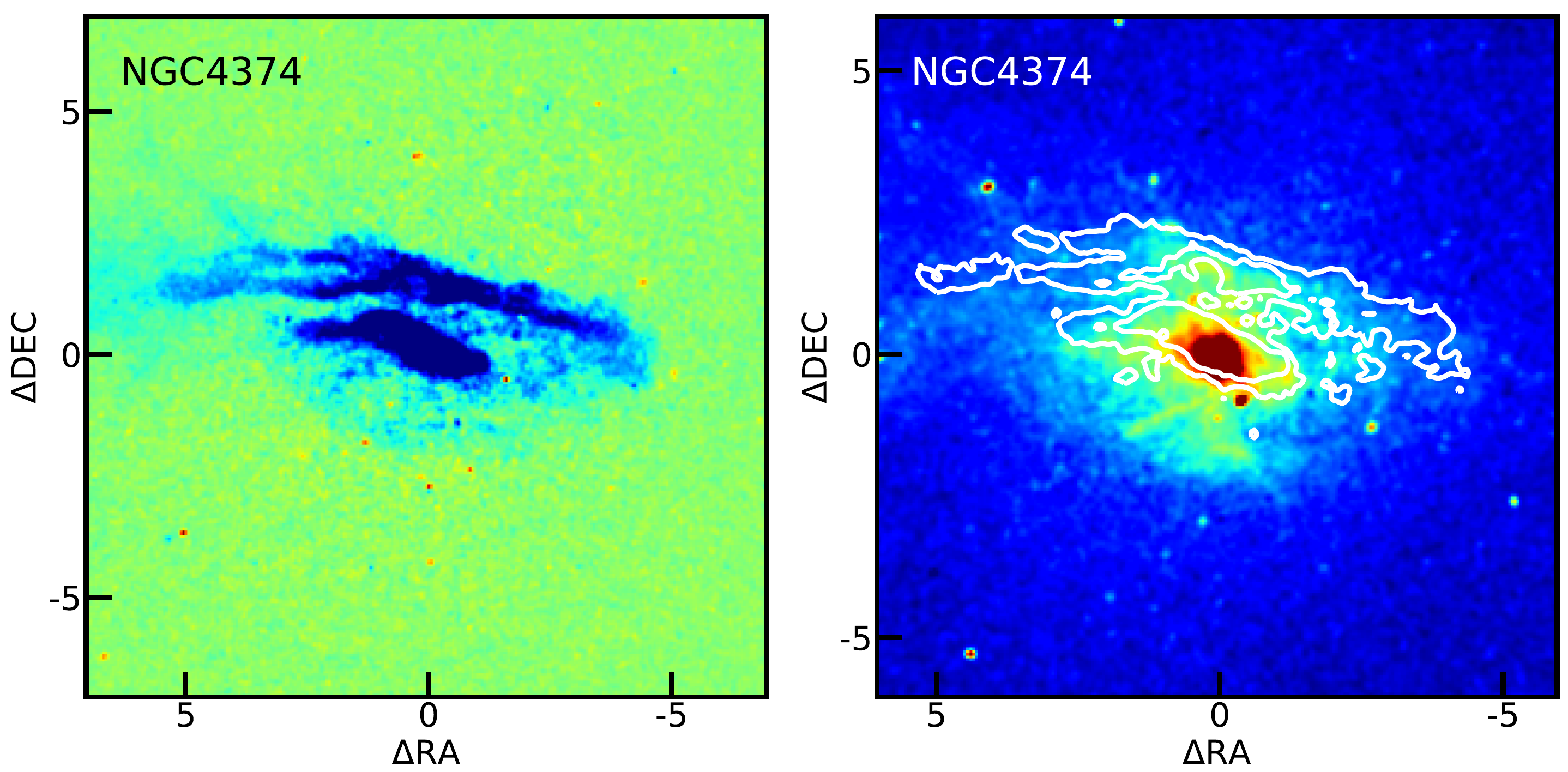}
  \caption{{\bf Left:} {\it HST} dust absorption map of NGC~4374.  The map
    appears in blue in this color scale, zero emission is green, and red area represent
    residual emission or emission bluer than the average stellar population in the galaxy. {\bf Right:} {\it HST} $H\alpha + [NII]$ map: a stellar emission continuum is removed from a narrow filter. The dust absorption is overlaid with white contours. }
\label{fg:4374_dust}
\end{figure}

 \begin{figure*}[ht!]
  \centering
  \includegraphics[width=12.5cm]{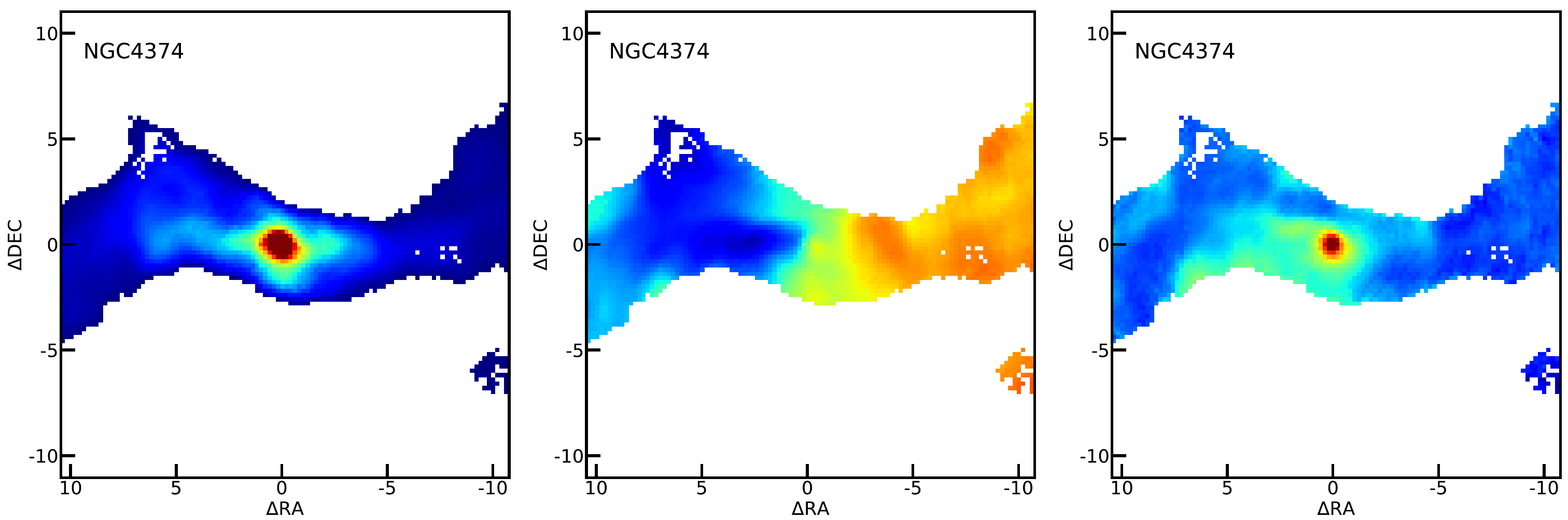}
  \caption{H$\alpha$ emission (left), velocity (center) and velocity
    dispersion (right) estimated on MUSE spectra for NGC~4374. On the color scale, the velocity ranges
    from -250 to 250 km/s, the velocity dispersion range from 0 to 250 km/s. The H$\alpha$ image is in units of 10$^{-20}$ erg cm$^{-2}$ s$^{-1}$.}
\label{fg:4374_ha}
\end{figure*}

SOFIA observations have not detected cool [CII] emitting gas in NGC~4374; the flux upper limit is reported in Table \ref{tab:CII_CO_gal}.
Although the [CII] line emission has not been detected, the galaxy
exhibits extended, stratified dust seen in absorption in the nuclear region with a significant cold dust emission observed in the far-IR that accounts for a total dust mass of 1.6$ \times 10^6 M_\odot$ (Figure \ref{fg:4374_dust}).
From {\it HST} data, 
NGC~4374 H$\alpha$+[NII] emission has a large number of features. A strong
central emission is extended in the East-West direction by about $0.5^{\prime \prime}$. Some antenna-like thin plumes point in the Southern direction
and extend about $1^{\prime \prime}$ from the center of the galaxy. Some very
large plumes extend more than $6^{\prime \prime}$ in the Eastern direction and
about $3^{\prime \prime}$ in the Western direction. Multiple H$\alpha+\rm [NII]$
small substructures are visible within $2^{\prime \prime}$ from the
center of the galaxy. The largest scale features of the H$\alpha$+[NII] emission
seem to correlate with the dust absorption as seen in Figure \ref{fg:4374_dust},
although it appears that the dust is systematically located about 0.5'' northward of the
 H$\alpha$+[NII] emission.
The velocity and velocity dispersion estimated with the MUSE spectra 
ranges from -250 to 250 km/s and from 0 to 250 km/s, respectively (Figure \ref{fg:4374_ha}).

CO(2-1) in NGC~4374 has been detected in the central $\sim 1^{\prime \prime}$ where the optical {\it HST} map shows a dusty disk with major axis misaligned by $\sim 90$ degrees with respect to the stellar one \citep{boizelle17}. 
NGC~4374 is classified as a slow rotator \citep{emsellem11} with almost no line of sight stellar rotation. Both the cold molecular gas and ionized gas kinematics show disk rotation that appear to be misaligned with respect to the stellar axis by $\sim 90 ^\circ$ \citep{boizelle17, walsh10}.

The X-ray morphology is strongly affected by AGN feedback.  It shows extended ``arms'' of gas compressed by AGN jet inflated radio lobes \citep[see also][]{finoguenov2001,finoguenov2008} (Figure \ref{fg:xrayflux}). 
 Radio emission observed at 1--2\,GHz exhibits strong radio mode feedback within the inner parts of the galactic atmosphere (Figure \ref{fg:vla_gas}; bottom left).

\vskip0.8cm
\subsection{NGC~4406}
\begin{figure}[hb!]
  \centering
  \includegraphics[width=8.5cm]{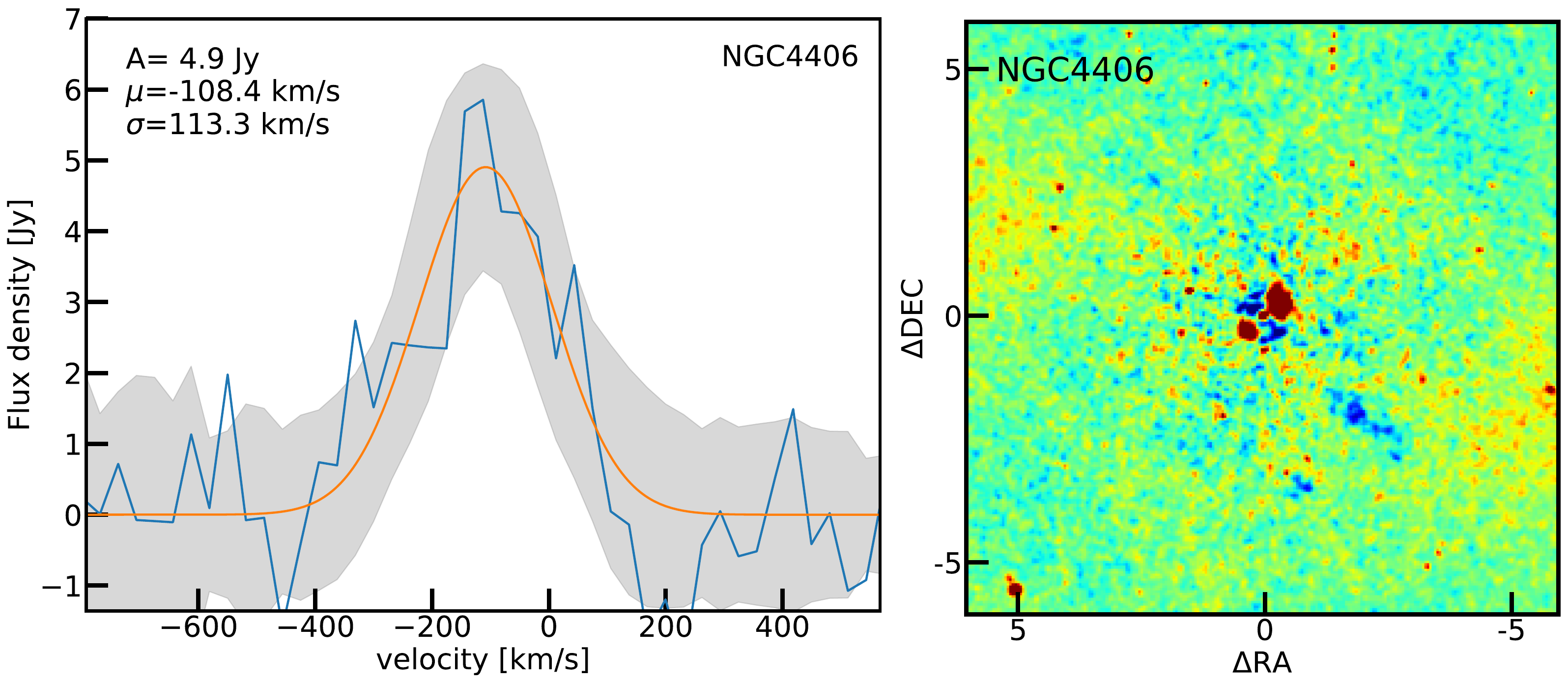}
  \caption{{\bf Left:} CII lines (blue solid lines) detected with SOFIA FIFI-LS in NGC~4406. The orange solid line
    represents a Gaussian fit to the line, the parameters of the Gaussian are indicated in top left corner (A: amplitude, $\mu$ :
    velocity center, $\sigma$ : velocity width), and the name of the galaxy in the top right corner. The grey shade area
    indicate the noise level in the spectrum. 
    {\bf Right:}
    {\it HST} dust absorption map. The dust absorption
    appears in blue in this color scale, zero emission is green, and red area represent
    residual emission or emission bluer than the average stellar population in the galaxy.}
\label{fg:4406_cii}
\end{figure}

NGC~4406 (M86) is a giant elliptical E3 (from RC3) galaxy located in
the Virgo cluster. 
SOFIA has detected cool [CII] emitting gas located within
the central 2 kpc. The detected gas is compact with a velocity dispersion of $\sim 115$ km\,s$^{-1}$ and is spatially
unresolved by the FIFI-LS instrument (Figure \ref{fg:4406_cii}).

Early IRAM single dish observations in the CO(1-0) molecular gas have resulted in upper limit flux with corresponding upper limit in $H_2$ molecular mass of $M_{\rm H_2}<  0.32 \times 10^8 M_\odot$ \citep{wiklind95}.
{\it HST} dust image shows some residual emission at its center and a plume
along the radial direction extending $\sim 4^{\prime \prime}$.
Morphologically complex warm ionized
gas has been detected through H$\alpha$ emission \citep{kenney08}.
linking NGC~4406 to the spiral galaxy NGC~4438, that lies $23^\prime$
(120 kpc) away.  
Using
{\it Herschel} far-infrared detectors, \cite{gomez10} detected several dust
clouds around NGC~4406 (total mass of $2-5 \times 10^6~M_\odot$), some
of which (most massive) are colocated with the H$\alpha$ emission
indicating the dust also comes from a collision with NGC~4438.
X-ray observations of NGC~4406
do not display obvious indications of AGN feedback (Figure \ref{fg:xrayflux}), but
the galaxy is particularly strongly affected by ram-pressure stripping
as it is supersonically falling into the Virgo cluster along a
direction close to our line of sight \citep{randall2008}. 
NGC 4406 shows at 1--2\,GHz only point-like radio emission (Figure \ref{fg:vla_gas}; top right).

\subsection{NGC~4552}
NGC~4552 (M89) is a giant elliptical E0 (from RC3) galaxy which is
falling into the Virgo cluster (it is 350 kpc to the east of M87, in
subcluster A). 
Cool [CII] emitting gas has not been detected in
NGC~4552 and the upper limit is reported in Table \ref{tab:CII_CO_gal}.

\begin{figure}[ht!]
  \centering
  \includegraphics[width=5.25cm]{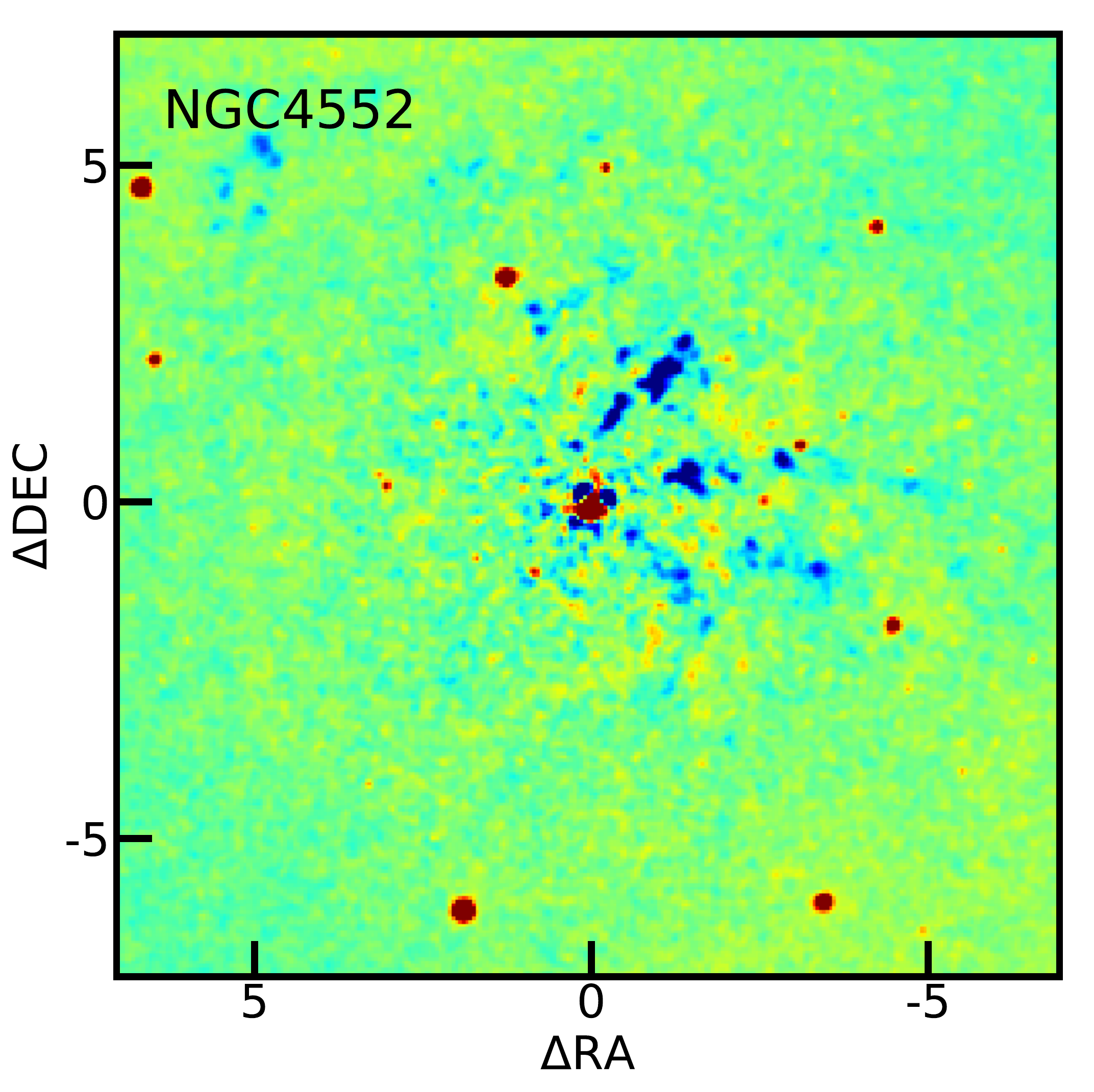}
  
  \caption{{\it HST} dust absorption map of NGC~4552.
   The dust absorption
    appears in blue in this color scale, zero emission is green, and red area represent
    residual emission or emission bluer than the average stellar population in the galaxy.}
\label{fg:4552_dust}
\end{figure}

The dust absorption map elaborated with {\it HST} data (see Figure \ref{fg:4552_dust}) 
accounts for minimal dust seen in absorption in the nuclear region.
There are hints of a few dust plumes along the radial direction extending out up to 4''/300 pc away from the galaxy center.
Low level detection at far-IR wavelengths accounts for a mere 3500~$M_\odot$ of total dust estimated with the spectral energy distribution (SED) modeling software, CIGALEMC \citep{amblard14}, 
although larger dust masses are reported in a recent paper by \cite{boselli21}.
Even if specific SED fitting codes could generate slightly different dust masses (similarly to \cite{boselli21}, in this work we have used results, among others, derived from CIGALE SED code (see section 4 and discussion in \citealt{amblard14})), we believe that most of the dust mass discrepancy is due to the evaluation of the diffuse and extended FIR emission that is given as input to the SED fitting process. There is indeed a significant difference between the two FIR sets of data (see Table \ref{tab:dustgals}, and \citealt{ciesla12,cortese14}). The discrepancy in the observed flux is puzzling, and can be ascribed to the procedures used to evaluate the radial extent of the diffuse low level emission. \\
CO has not been detected in NGC~4552: IRAM observations only provide a CO(1-0) flux upper limit (Table \ref{tab:CII_CO_gal}).
H$\alpha$+[NII] emission estimated from {\it HST} data did not return any
significant signal.
Early ground-based observations have reported H$\alpha$+[NII] emission in the central regions \citep{trinchieri91, macchetto96}. 
More recent very deep narrow-band H$\alpha$+[NII] imaging carried out with MegaCam at the CFHT shows a diffuse and filamentary distribution of the ionised gas \citep{boselli21}.
\ X-ray emission is detected at the center of the galaxy with a sharp edge 3.1 kpc to the north,
a cool ($kT = 0.51^{+0.09}_{-0.06}$ keV) tail extending 10 kpc to the south
of the galaxy, and two 3 kpc-long ''horns'' extending south on the sides
of the north edge (Figure \ref{fg:xrayflux}). 
The X-ray morphology is strongly affected by
AGN feedback: the AGN shock
induced ring and the tail are interpreted as characteristic features of a
supersonic ram-pressure stripping of the gas, produced by the motion
of NGC~4552 in the Virgo ICM and estimated an infall velocity of 1680 km~s$^{-1}$
\citep{machacek06a, machacek06b}.  
Figure\,\ref{fg:vla_nogas} (top image) shows the VLA radio data at 1--2\,GHz. 
The galaxy has a radio source with jets extending to $\sim 10$ kpc along the north-south direction where the radio lobes coincide with cavities in the X-ray images \citep{shurkin08}.

\subsection{NGC~4636}

As mentioned before, NGC~4636 
is not a member of the original SOFIA [CII] galaxy sample.
It is part of the galaxy sample observed in the [CII]
line by the {\it Herschel} observatory  \citep{werner14}, and here it serves as a term of comparison
for the interpretation and understanding of feedback processes acting on the galaxies under investigation.\\
Cold gas in the group-centered elliptical galaxy NGC~4636 has been reported in recent publications \citep{werner14, temi18}.
A map of the integrated [CII] line flux
obtained with {\it Herschel} PACS shows extended
emission, presumably in filamentary structures, in the central $\sim$3 kpc with a velocity dispersion of $\sim$350 km/s \citep{werner14}.
The extended [CII] line emission is cospatial with near-infrared $H_2$ emission and filamentary structures in the H$\alpha$ + [NII] nebula. CO(2-1) molecular gas is detected in the form of off-center orbiting clouds distributed in the center of the galaxy.
The new ALMA CO(3-2) data confirm the presence of a compact molecular cloud few arcseconds off the center. The associated total molecular mass is estimated to be $2.6 \times 10^5 M_\odot$.
Small, chaotically arranged dusty fragments and filamentary structures are visible in {\it HST} dust absorption maps \citep{temi18}. 
The H$\alpha$ + [NII] emission is extended, but remains relatively compact concentrated in the innermost $r \le 1.5$ kpc of the galaxy with the peak centered in the optical galactic nucleus. Filamentary structures and bright
 compact knots are apparent in the central few kpc of the galaxy.
 NGC~4636 exhibits strong radio mode feedback within the
inner parts of the atmosphere. 
 Radio observations show AGN jets with energy deposited just outside a central, compact, X-ray bright, gaseous core of NGC~4636 (Figure\,\ref{fg:vla_gas}; bottom right).
The X-ray gas morphology appears affected by AGN feedback, showing ``arms'' of gas compressed by AGN jet inflated radio lobes \citep[see also][]{baldi2009,finoguenov2008}.

\subsection{NGC~4649}
\begin{figure}[ht!]
  \centering
  \includegraphics[width=5.25cm]{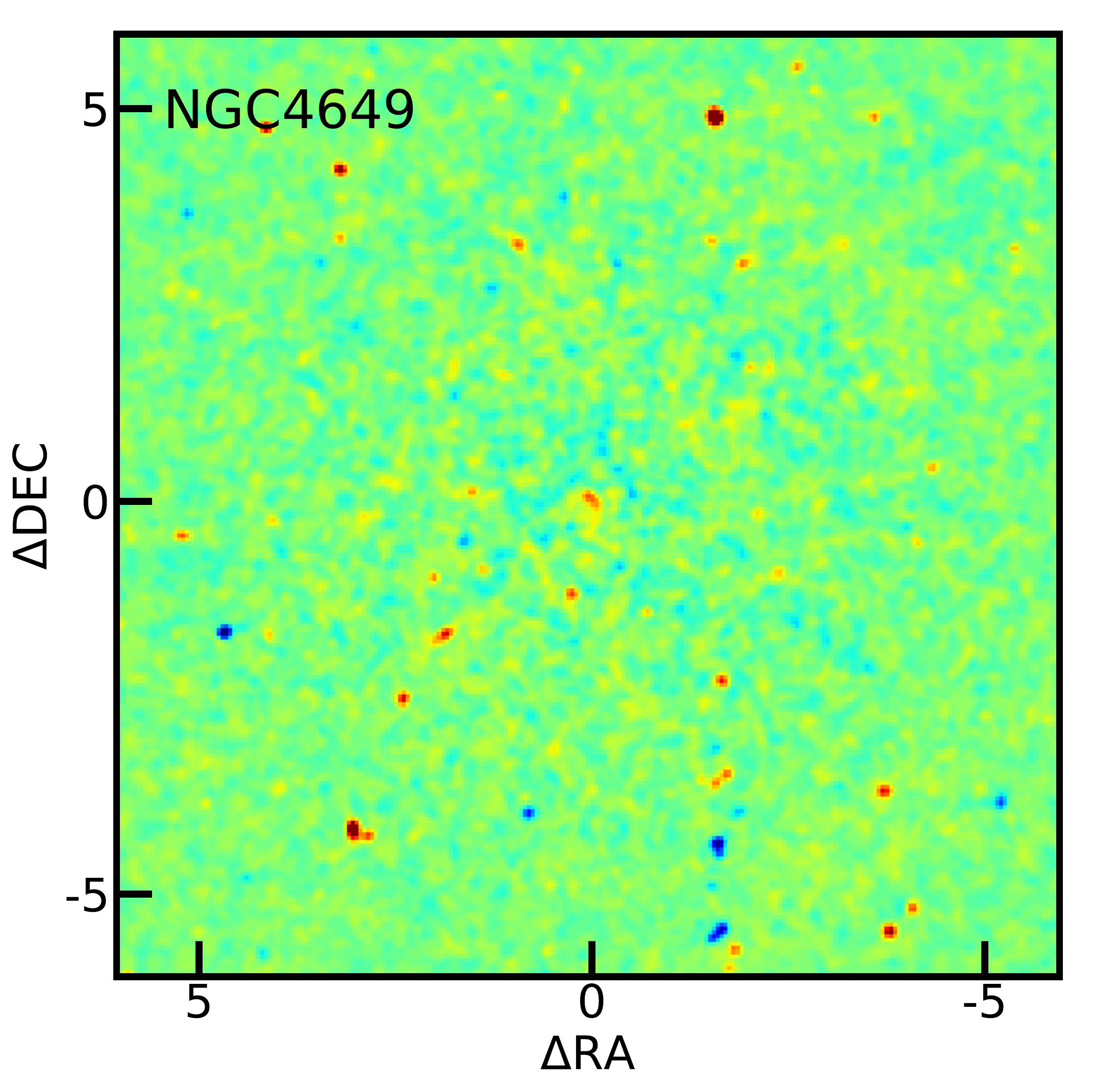}
  
  \caption{{\it HST} dust absorption map of NGC~4649.
   There is no evidence of dust in absorption in the central $\sim 1.2$ kpc.}
\label{fg:4649_dust}
\end{figure}

NGC~4649 (M60) is a giant elliptical E2 (from RC3) galaxy at the
eastern edge of the Virgo cluster and has a neighbouring spiral Sc galaxy NGC~4647
$\sim 2^\prime.5$  away. NGC~4649 appears to be a truly cold-gas-free galaxy: 
i) SOFIA FIFI-LS observations, although not very deep, did not detect [CII] emitting gas, 
ii) dust absorption map elaborated with {\it HST} (see Figure \ref{fg:4649_dust}) shows no evidence 
of dust in the central kpc. Also, far-IR cold dust emission has not been detected in this galaxy - early ISO null observations \citep{temi04} have been confirmed at 160~$\mu$m  and 250~$\mu$m with an estimated total amount of dust of the order of 300~$M_\odot$, iii) recent ALMA observations  do not reveal any CO molecular gas in the central $\sim 30^{\prime \prime}$.

\begin{figure*}[hb!]
  \centering
  \includegraphics[width=14.5cm]{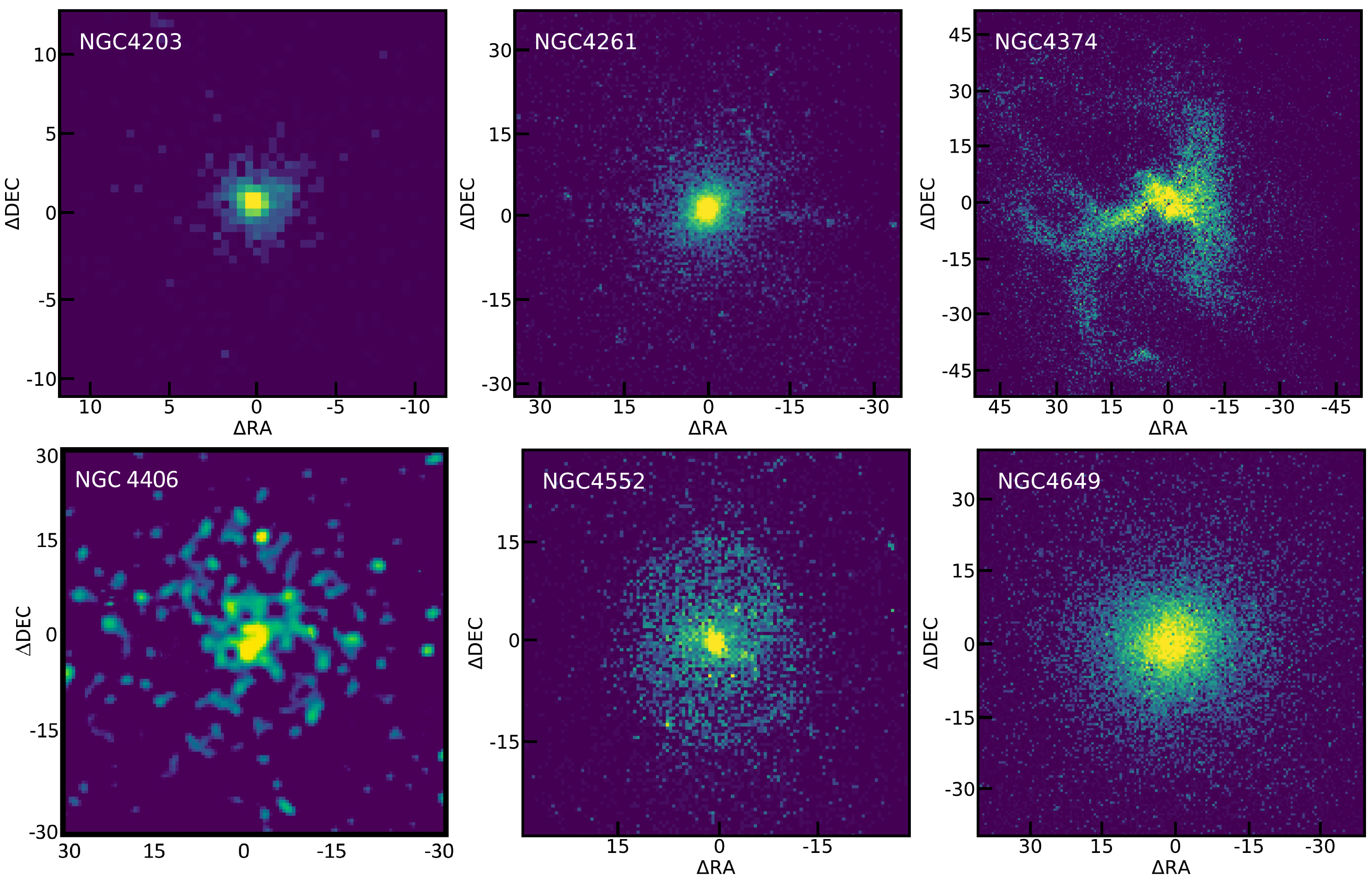}
  \caption{{\bf X-ray flux images in the 0.5-2 keV band of the galaxy sample}
      }
  \label{fg:xrayflux}
\end{figure*}

 \cite{trinchieri91} have reported H$\alpha+\rm [NII]$ emission  in the central region of NGC~4649, but more recent 
measurements with APO/SOAR do not confirm detection.
The X-ray gas morphology in NGC~4649  (Figure \ref{fg:xrayflux}) appears spherically symmetric and apparently undisturbed by AGN energy release.
In relation to the reported faint nuclear radio source, 
small disturbances in the X-ray-emitting gas \citep{shurkin08,dunn10, humphrey13} have been reported on $<3$ kpc scales, corresponding to $\sim 40 ^{\prime \prime}$.
 \cite{shurkin08} showed X-ray cavities at the location of radio lobes (northeast and southwest) that extend $\sim 20^{\prime \prime}$. 
 Using deep X-ray images, \cite{humphrey08} and \cite{paggi17} did not find convincing evidence of any morphological disturbances, reporting  a relaxed and symmetric X-ray morphology at scales from 3  to 12 kpc.  
 On large scales ($ >160^{\prime \prime}$, 12 kpc),
\cite{wood17} showed some evidence of ram pressure stripping (edge in the
surface brightness and wing-like structures)
due to the motion of the galaxy in Virgo
ICM.
\cite{lanz13,dabrusco14}
suggested tidal interaction between NGC~4649 and NGC~4647 based on the
shape of their spectral energy distributions and the spatial
distribution of LMXBs.
Figure \ref{fg:vla_nogas} (bottom image) presents the VLA radio data at 1--2\,GHz showing a weak radio source with small-scale jets extending to $\sim 6$ kpc along the north-south direction where the radio lobes 
coincides with cavities in the X-ray.

\begin{figure*}[hb!]
  \centering
  \includegraphics[width=14.5cm]{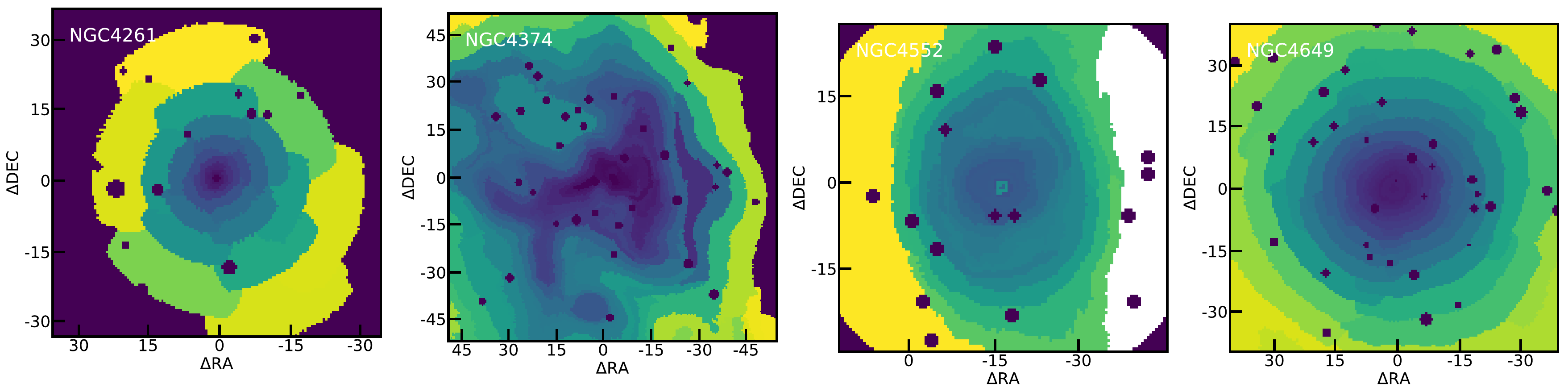}
  \caption{X-ray entropy of  NGC 4261, 4374, 4552, 4649}
  \label{fg:xrayent}
\end{figure*}

 \begin{figure*}[ht]
  \centering
  \includegraphics[width=14.5cm]{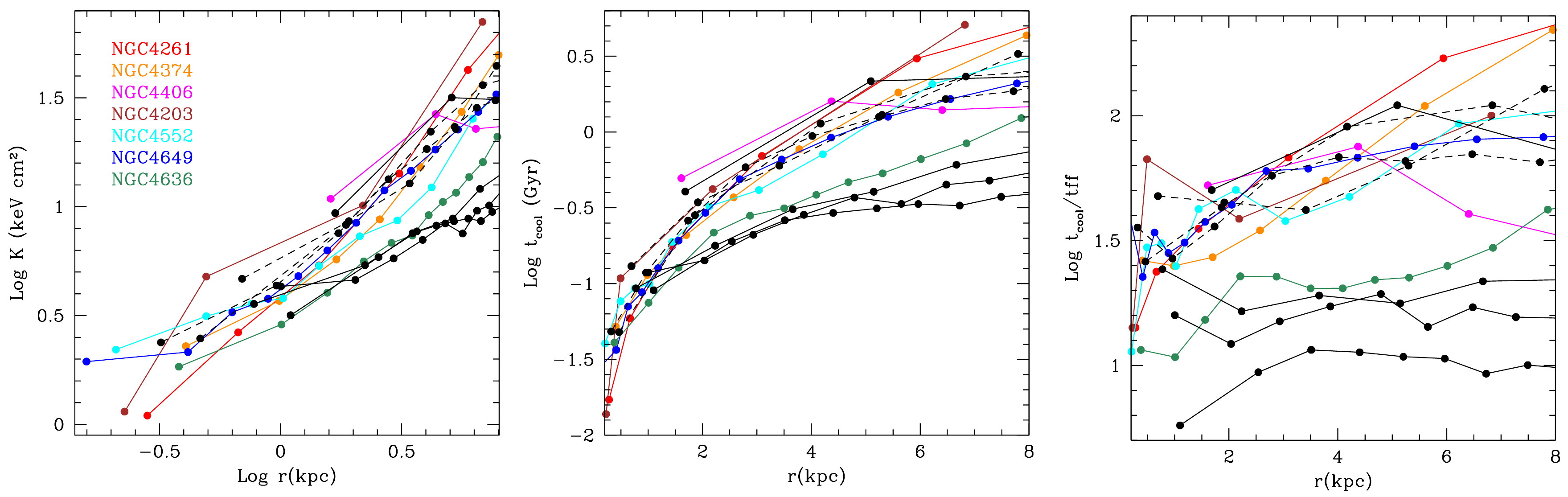}
  
  \caption{Deprojected entropy K, $t_{\rm cool}$, and $t_{\rm cool}/t_{\rm ff}$ ratio radial profiles. In color solid lines refer to galaxies in our sample. Black lines identify galaxies with (solid) and without (dashed) extended cool gas in a parent sample studied by \citet{werner14}. The solid green line refers to NGC~4636, the galaxy with extended and filamentary cool gas  that is in common between the two samples.
  }
  \label{fg:thermo}
\end{figure*}

\begin{table*}[ht!]
\footnotesize
\caption{Summary of the multiphase gas properties.}
  \label{tab:sum}
\begin{center}
  \begin{tabular}{lllllccc}
Galaxy  &      Cold Gas   &  H$\alpha +\rm [NII]$ Gas     & X-ray Gas &  Radio &   $P_{jet}$\tablenotemark{a} & $P_{\rm e}$ \tablenotemark{b}& K \tablenotemark{c}   \\ 
                  & Reservoir   & Morphology     & Morphology       & Morphology   &($10^{42} erg~s^{-1})$ & ($\rm erg~cm^3$) & ($\rm keV~cm^2$) \\ 

\hline
 NGC~4203   &  Yes, detected in  &  Central, &  Affected by ram     &  Faint point-like radio source    & \nodata  &$4.7\pm 1.6$  & $10\pm 2$  \\
                      &     [CII], CO, Dust   &  compact  &  pressure stripping              &    &&  &  \\
NGC~4261   &  Yes, detected in &  Central, &  Relaxed; spherically     & Large power, jets penetrating    &  $21.5\pm 4.3$ & $27.0\pm 0.2$ & $7.5\pm 0.3$   \\
                     &    [CII], CO, Dust      &   compact disc  &   symmetric        & the atmosphere and depositing         &  &&     \\
  &    &   &     &    energy at large radii     &   &    & \\
NGC~4374  & Yes, detected in & Extended   &  AGN disturbed,     &    Strong radio mode feedback     & $15.3 \pm 4.6$ & $24.5\pm 0.2$   &$8.5\pm 0.4$    \\
                     &   CO, Dust          &                     &   arms  of compressed  &     within the inner parts of the   &  &     &\\
                     &                            &                     &     gas                               &  atmosphere   &   &   & \\
NGC~4406   &  Yes  &    &   Affected by ram  &   Faint point-like radio source      & \nodata  & $2.0\pm 0.1$   &$15\pm 1$    \\
                    &       &    &   pressure stripping   &        &    &  &  \\
NGC~4552 & No \tablenotemark{d}  &    &   AGN disturbed,  &  Strong radio mode feedback      & $1.6 \pm 0.4$ &    $21.3\pm 0.5$   &$7.3\pm 0.3$   \\
                    &       &    &   ring and tail   &    within the inner parts  of the   & &  &   \\
                   &       &    &                           &  atmosphere    &  & &    \\
NGC~4649   & No   &    &  Relaxed;    &  Small lobes in the  central  &   $1.3 \pm 0.4$   & $72\pm 10$   &$8.5\pm 0.1$ \\
                   &        &     &  spherically symmetric      &   region  &   &   & \\
NGC~4636  &  Yes, detected in  &  Extended;  &  AGN disturbed,          &  Strong radio mode feedback  & $0.30 \pm0.08$ & $18.3\pm 0.4$ &$5.6\pm 0.2$   \\
                    &  [CII], CO, Dust       &  Filamentary& arms of compressed  &          within the inner parts of the   &    & &   \\
                    &                                &                      &  gas                              &          atmosphere                         &     &    &   \\
\hline
 \end{tabular}
  \tablecomments{
  $^{\rm a}$ The mechanical power output of the AGN (including both jets) is taken from  from \citet{allen06,shurkin08,kolokythas18}; 
{$^{\rm b}$ Central pressure at 0.5 kpc; $^{\rm c}$ Entropy at 2 kpc; $^{\rm d}$ See discussion in section 5.1.2}
 }
  \end{center}  
\end{table*}

\section{Discussion}
Given the large multiwavelength dataset available for the galaxy sample presented here, we wish
to discuss in detail the content, kinematics and spatial distribution of the observed multiphase gas.
Although small, our galaxy sample shows a remarkable diversity of systems 
with a substantial range of 
cold/warm gas content, distribution and kinematics. 
Thus, in combination with the thermodynamical properties of the hot gas, and the analysis of the radio properties in terms of morphology and energy power deposition, we wish to establish a path for the evolution of the gas in these galaxies that reflects their current physical condition. 

SOFIA FIFI-LS observations reveal [CII] line emission in 3 (NGC~4203, NGC~4261 and NGC~4406) out of 6 observed galaxies. Multiwavelength data for NGC~4649, one of the three galaxies undetected in [CII], do not show the presence of any other cool or cold gas phases or significant amount of dust. 
NGC~4552 eluded [CII] detection and shows inconsistent  data regarding its cold gas content: in addition to [CII], it has not been detected in CO and shows large discrepancy in the amount of dust accounted via SED fitting procedures.
On the other hand, NGC~4374 (M84) displays filamentary H$\alpha$+[NII] emission, dust, and molecular gas. 
The non-detection of [CII] emission in this system is therefore all the more puzzling. 
In our galaxies (except for NGC 4636) the distribution of the detected cool gas phases is centrally concentrated and fairly compact. The molecular gas spans a range of masses 
from $6.1 \times 10^6$ $M_\odot$ in NGC~4374 to $2.3 \times 10^7$  $M_\odot$ in NGC~4203. Both in the cool-gas-rich and cool gas poor galaxies (except M86, which might be an outlier due to its supersonic merger through the Virgo cluster) the X-ray emitting gas has relatively steep entropy profiles and the radio jets are on, depositing their energy in the hot atmospheres of the galaxies.\\

\subsection{Origin of the Cold Gas} \label{sec:origin}
The origin of the cold gas in massive ellipticals
is still a matter of debate. 
While minor mergers may be responsible for a significant fraction of the cold gas mass in low to intermediate mass early-type galaxies  \citep[e.g.][]{davis11}, internal sources
of neutral/molecular material, like hot gas cooling and stellar mass loss, are likely dominating the cold gas budget in massive ellipticals \citep{davis11,lagos14,david14,werner14,valentini15,sheen16,gaspari17_rain,gaspari18,babyk19}.
Moreover, the distribution and dynamics of the warm ($T\sim 10^4$ K) ionized gas can provide
additional clues to the origin of the multiphase gas.
Table \ref{tab:sum} gives a qualitative summary of the multiphase gas status  in the investigated galaxies.

Figure \ref{fg:thermo} shows the deprojected entropy profiles of the hot X-ray emitting gas of the individual galaxies along with the cooling time $t_{\rm cool}$ and the $t_{\rm cool}/t_{\rm ff}$ ratio, where $t_{\rm ff}$ is the free-fall time. The method used to derive the profiles is presented in detail in \citet{lakhchaura2018}. 
In color solid lines refer to the radial profiles of our galaxy sample.
Black lines identify galaxies from the sample of group-centered ellipticals studied by \citet{werner14}.
Their investigation reveals a clear separation in the radial profiles of thermodynamic properties of the hot gas between the galaxies that exhibit  extended filamentary structures in the cool gas phase and the galaxies without cool gas. Such a distinction is apparent in 
Figure \ref{fg:thermo} where the solid black lines refer to galaxies with extended cool gas and the dashed black lines identify galaxies almost free of cool gas.
NGC~4636 (green solid line), the galaxy with extended and filamentary cool gas  that is in common between the two samples, traces the upper envelope of the solid black lines.
The dichotomy in the entropy profiles in \citet{werner14} supports the scenario in which the cold gas originates from the cooling of the hot gas phase: the condition of low core entropy in such cold gas-rich galaxies promotes non-linear density perturbations able to trigger very frequent and extended top-down cooling
\citep{gaspari12_ETGs, gaspari13, sharma12, werner14, brighenti15,valentini15, voit15, li15}. 
On the other hand, our sample is likely covering the second phase of the internal self-regulation cycle, in which the halo tends to be overheated and in the more quiescent rotating stage.
Indeed, with the exception of NGC~4636, the other galaxies in our sample with CO emission (Table \ref{tab:CII_CO_gal}) show molecular gas is in the form of compact clouds or, as in NGC~4261 and NGC~4374, in a compact rotating disc. Similarly the [CII] emitting gas is observed in compact regions near the center, with the warm H$\alpha+\rm [NII]$ gas being more extended.
This differs from several cold gas-rich galaxies studied by \citet{werner14}.
Fig.~\ref{fg:thermo} further shows that their entropy and cooling time radial profiles tend to overlap with the profiles identified by galaxies with low or no cool gas in the \citet{werner14} sample, albeit the scatter is large, above 0.5 dex.
Our sample does not show a strong separation in entropy or cooling time profiles among galaxies with or without nuclear cool gas, suggesting that such galaxies are in a different mode of the condensation cycle, rather than in the more extended precipitation stage.
We will discuss in Sec.~\ref{s:cond} other key diagnostics of the condensation scenario; in particular, the $C$-ratio\,$\equiv t_{\rm cool}/t_{\rm eddy}$ (unlike the $t_{\rm cool}/t_{\rm ff}$) is better suited to unveil the different condensation regimes between central and non-central galaxies, with the the latter expected to develop a nuclear rain rather than extended filaments.
The amount of molecular gas mass in the two parent samples shows another difference with Werner's sample.
Table \ref{tab:CII_CO_gal} reports the molecular gas mass based on the conversion from the CO flux. 
The molecular mass is computed by using (\citealt{bolatto13}) 

\begin{equation}
M_{\rm mol} = 1.05 \times 10^4 \left(\frac{X_{CO}}{2 \times 10^{20}}\right) \; \frac{S_{CO} \Delta\nu \; D_L^2}{ (1+z)} \; M_\odot,
\label{eq:mmol}
\end{equation}
where $S_{CO} \Delta\nu $ is the integrated line flux density in Jy\,km\,s$^{-1}$
in the ground rotational transition $J = 1 \to 0$, $D_L$ is the luminosity distance to the source in Mpc, and {\it z} is the redshift. 
We use the {\it reference} conversion factor $X_{CO}=2\times 10^{20}$ cm$^{-2}$ (K km s$^{-1}$)$^{-1}$
 to evaluate the molecular gas mass in each galaxy.
When detected, the mass of cool gas phase in our galaxy sample  
is often larger than the total mass of molecular gas detected in the central group galaxies that exhibit diffuse and filamentary cool gas extending to several kpc \citep{werner14, temi18}.
For instance, the group-centered elliptical galaxies, NGC~5846, NGC~4636, and NGC~5044 have extended and filamentary H$\alpha$ emission that is well matched by the 
gas emitting the [CII] line. They have molecular gas in the form of off-center orbiting clouds with associated total molecular mass from $2.6 \times 10^5 M_\odot$ in NGC~4636 to 
$6.1 \times 10^7 M_\odot$  in NGC~5044 over a region extending over few kpc \citep{david14,temi18}.
In contrast, the three sample galaxy NGC~4261, NGC~4203, and NGC4374 have molecular gas masses of $1.8 \times 10^7 M_\odot$ within $\sim 200 - 300$ pc, $2.3 \times 10^7 M_\odot$ within $\sim 300$ pc, and $6.1 \times 10^6 M_\odot$ within $\sim 200 $pc
\citep{boizelle17}, 
respectively.

\begin{figure}
  \centering
  \includegraphics[width=8.8cm]{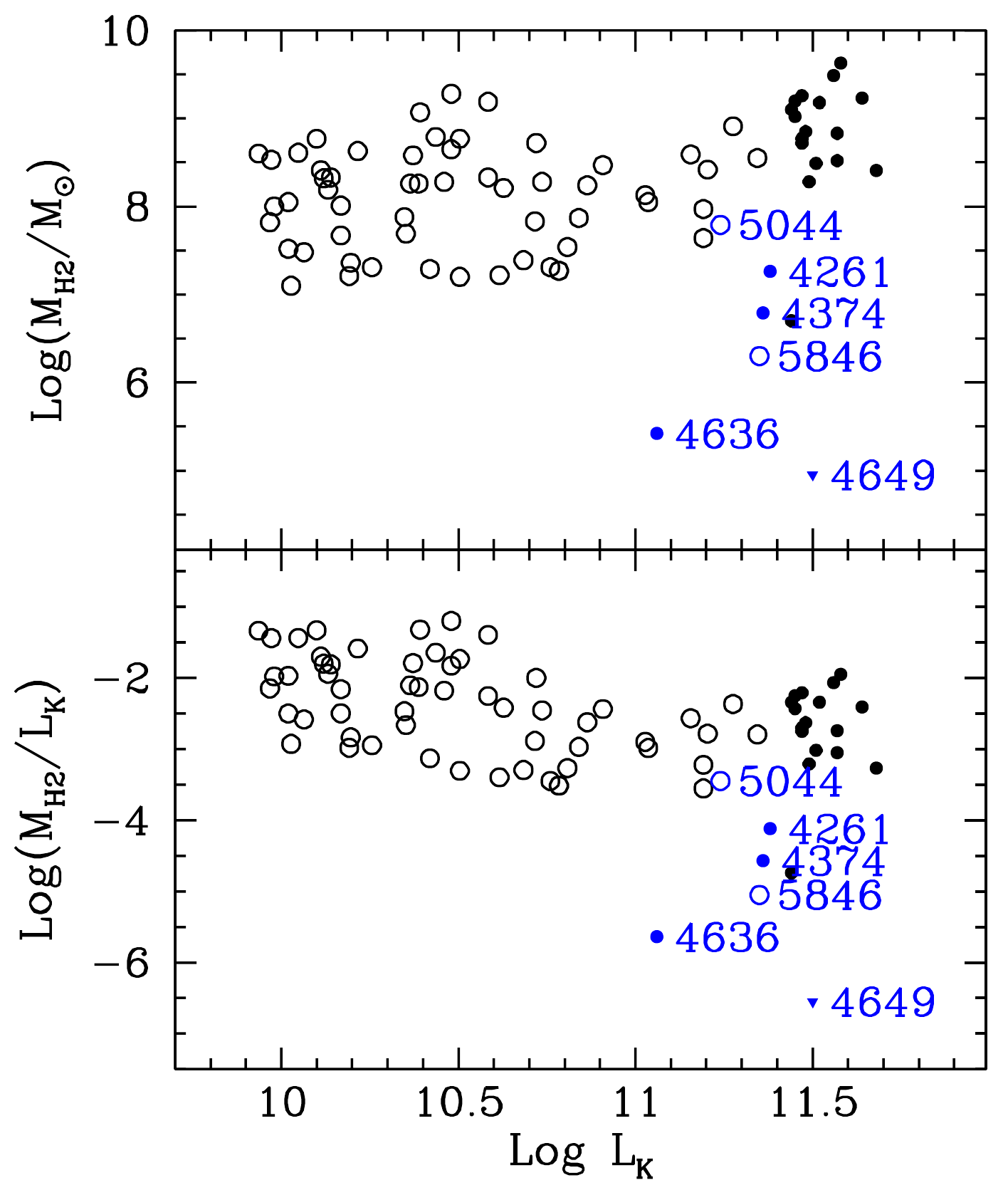}
  \caption{
  Molecular gas mass(top panel)and molecular gas mass to stellar luminosity ratio (bottom panel) plotted against Ks-band luminosity. 
  Black circles refer to MASSIVE (solid circles)  and ATLAS3D (open circle) CO detected survey galaxies (from \cite{davis19} and \cite{young11}. Blue circles represent ALMA CO detected galaxies of our sample (solid symbols) and the group centered galaxies in \cite{werner14} (open symbols). The CO undetected galaxy NGC~4649 is shown as an upper limit by a solid blue triangle.}
  \vspace{+0.1cm}
  \label{fg:H2}
\end{figure}

It is worth noting that, even considering the large range in molecular gas mass of the galaxies discussed here, they all show very low values of H$_2$ mass when compared to a large survey of early-type galaxies. Figure \ref{fg:H2} shows the molecular gas mass to stellar luminosity ratio plotted against Ks-band luminosity.
The blue symbols identify galaxies in our sample and the parent sample of \cite{werner14} of the objects that have been detected in CO with ALMA \citep[]{temi18}.
It is known that ALMA observations of CO emitting gas in these galaxies are not accounting for the large scale diffuse CO emission, if present. In group centered galaxies like NGC~5044 and NGC~5846 the ALMA integrated flux density is only $\sim$ 20\% of that in the IRAM 30m data, due to the presence of diffuse emission that is resolved out in the ALMA data \citep[e.g.][]{david14, temi18}.
Nonetheless, when compared to large early-type galaxy surveys, their detected CO fluxes account for much lower H$_2$ masses. ALMA deeper observations allow to explore for the first time the parameter space occupied by the galaxies investigated here.

\vskip0.3cm
Multiple sources may be at play to account for the cold gas present in our galaxy sample. Driven by the results from the extensive data available,
in the next sections, we examine and evaluate possible scenarios for the 
%formation 
origin of the cold gas in our sample.

\subsubsection{External origin of the cold gas} \label{s:external}

The reported results show some similarity with the findings of a study of a sample of Low Excitation Radio Galaxies (LERG) 
(i.e.~giant elliptical galaxies) 
by \citet{ruffa19a,ruffa19b}. The analysis of the molecular gas component in a volume-limited sample of LERGS shows that rotating (sub-)kpc molecular disks are common in LERGs. The bulk of the gas in the disks appears to be in ordered rotation and is co-spatial with dust lanes and disks of dust. For several targets, they also find evidence
in support of a possible external origin of the gas 
indicating that the environment in which the galaxies reside have significant impact on the generation of their cool gas. 
On the other hand, other observed samples of massive galaxies with rotating gas/disks show evidence of internal condensation (e.g., \citealt{juranova19,juranova20}).
In our sample, rotating (sub-)kpc gas disks, revealed in CO emitting gas and dust extinction maps, are observed in three out of five systems that have a cold gas reservoir (see \S\ref{sec:results}).

It is very unlikely that group-centered elliptical galaxies, like the galaxies probed by \citet{werner14,david14,temi18}, have acquired their cold gas via recent mergers with dust-rich galaxies. In  most cases, this is corroborated by several lines of evidence, like the lack of any optical evidence for recent mergers, the
mean stellar ages that are too old to be consistent with a recent merger, or the FIR emission that does not correlate with $H_\beta$ stellar ages \citep{temi05}.
In contrast, galaxies in our sample live in more diverse environments characterized by lower galaxy density, small groups and pairs such that it is conceivable that  they may have acquired at least part of their gas externally via galaxy-galaxy interactions \citep[e.g.][]{sabater13}. Indeed lines of support include:
 {\it i}) recent observations of H$\alpha$ emission in NGC~4406 \citep{kenney08}, some of which is
colocated in the region of its X-ray plume, suggest a collision with
NGC~4438 as its origin; {\it ii}) NGC~4261 has a well defined disk of dust, 240 pc in diameter,
that is aligned with its radio axis and perpendicular to the stellar
component of the galaxy;
{\it iii)} in NGC~4374 the dust lane major axis is strongly misaligned with the major axis of the stars and the kinematic of molecular gas is inconsistent with a stellar mass loss origin;
{\it iv}) the giant elliptical NGC~4552 which is
falling into the Virgo cluster, has central X-ray emission features \citep{machacek06a}
characteristic of a supersonic ram-pressure stripping of the gas, produced by its motion in the Virgo ICM.
Overall, albeit suggestive, these remain qualitative features that require to be further tested with future targeted and deeper observations. 
\\

 \begin{figure}
  \centering
  \includegraphics[width=8.5cm]{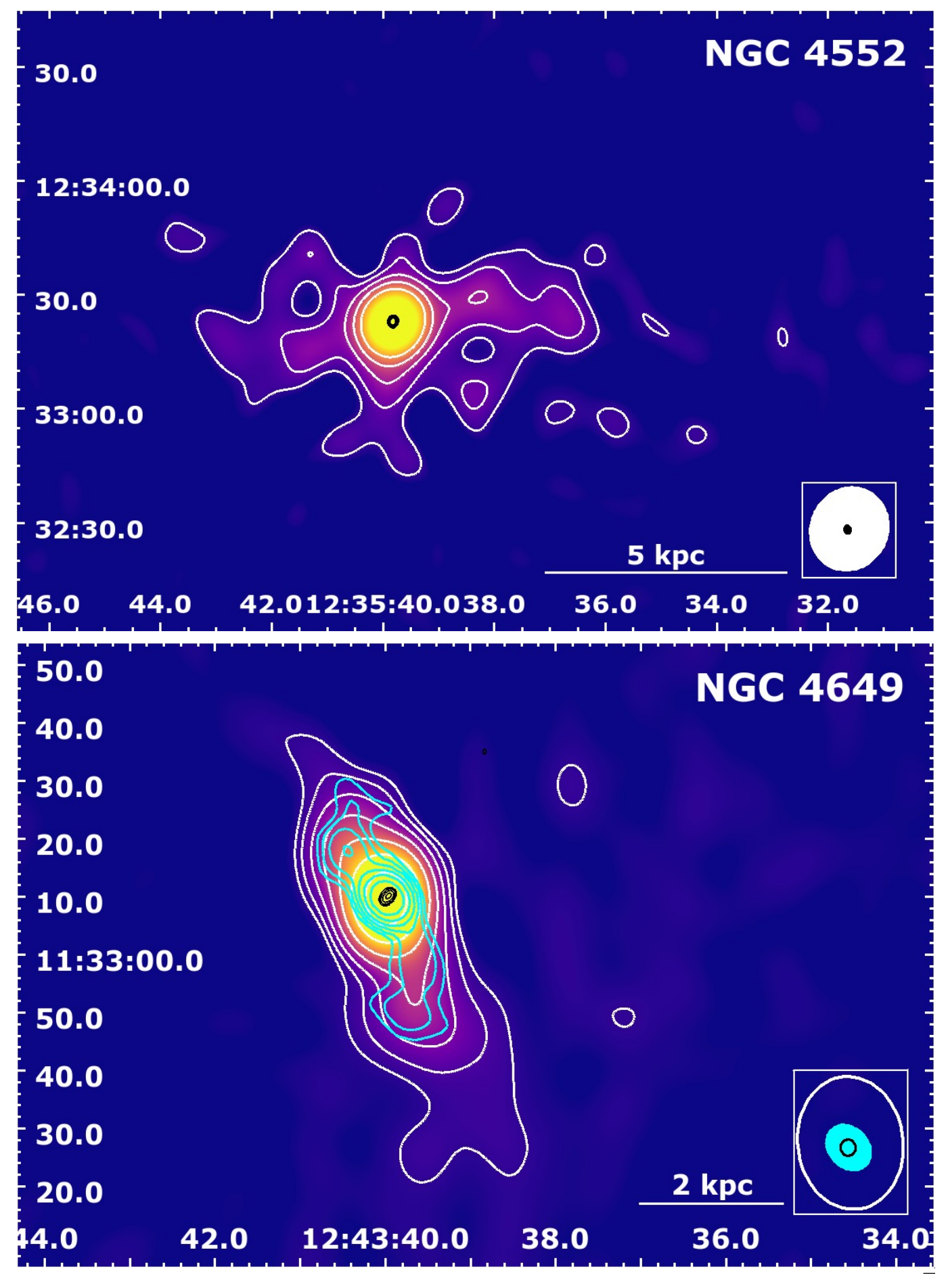}
  \caption{VLA radio total intensity maps at 1--2\,GHz for NGC\,4552 ({\bf top}) and NGC\,4649 ({\bf bottom}) in the C and D configuration, respectively. The maps are overlaid by the white contours of these compact configurations, the black VLA A configuration and cyan VLA B configuration contours \citep[previously published by][]{dunn10}. In all cases, the contours are created at [1,2,4,8,16,...]$\times\,5$ RMS noise and the restoring beams (resolution) are shown in the bottom right corner of all radio maps (for more details see Table\,\ref{tab:radio_reduction_details}).}
  \vspace{+0.1cm}
  \label{fg:vla_nogas}
\end{figure}

\subsubsection{Top-down condensation}\label{s:cond}
Considering that our sample galaxies harbor a supermassive black hole (SMBH) at the center, it is expected that AGN feeding and feedback still plays an important role in shaping and regulating the cool gas content in these systems. 

In the AGN self-regulation cycle the feedback and accretion on to the SMBH are effectively linked over nine orders of magnitude (\citealt{gaspari20} for a recent review): feeding via cold cloud condensation (known as Chaotic Cold Accretion -- CCA; \citealt{gaspari13}) triggers kinetic feedback in the form of jets and outflows, which can re-heat the hot halo and later stimulate thermal instability via turbulence (\citealt{yang19,wittor20}), further inducing filamentary condensation and accretion. This creates a self-regulated duty cycle, which alternates between periods of more intense CCA rain and more quiescent over-heated periods. 

In the following section, we compare the multiphase gas observations for our galaxy sample with key features as predicted by such AGN feeding and feedback scenario at different stages of the cycle, to test consistency with the top-down condensation.\\

\noindent
{\bf Cool gas-free galaxies} --  NGC~4649 shows no evidence of cool gas. The galaxy eluded detection in CO and [CII] lines and in dust emission via deep Spitzer observations. Warm H$\alpha+\rm [NII]$ emitting gas has not been securely detected either. 
As mentioned in section 5, the cold gas content in NGC~4552 shows some inconsistency.  On one hand the non-detections in the [CII] emitting gas as well as in the CO line indicate that the cold gas reservoir, if present, it is not large. On the other hand, because of the non-detection in the CO line emission, \citet{boselli21} estimate a mass of molecular gas Log M$_{H_2} = 8.63 \rm \ M_ \odot$ derived from the total dust mass and a gas-to-dust ratio of 80, rather than using the canonical conversion from equation \ref{eq:mmol}. We regard NGC~4552 as a system with low cold gas content, however given the inconsistent data available on the derived molecular gas mass, a note of caution is advised. 
Figure \ref{fg:vla_nogas} shows VLA radio data at 1--2\,GHz. Both galaxies host a weak radio source with small-scale jets extending to $\sim 10$ kpc in NGC~4552 and $\sim 6$ kpc in NGC~4649.
The symmetric X-ray gas morphology, the lack of cold gas reservoir in conjunction with a very week AGN jet may be indicative of a specific stage in the AGN self-regulation cycle. 
This is indeed the AGN starvation phase caused by the lack of the cold gas fuel: cool gas, previously condensed from the hot phase via cooling instabilities, is heated and eventually destroyed past the intense AGN feedback episode where large kinetic energy  is released  
to the surrounding medium. The injected energy drastically reduces the cold gas fuel in the nuclear region which, in turn, reduces the SMBH accretion rate and eventually brings the jet activity to a halt (e.g., \citealt{gaspari12_ETGs}).

NGC~4649 is approaching this phase where the AGN feedback energy release is drastically decreasing due to the lack of cold gas reservoir. This is also consistent with the current thermodynamic properties of the hot gas. The central gas pressure is the highest among all the galaxies in the sample (see Table \ref{tab:sum}).
In addition, the hot gas pressure and entropy maps are regular, smooth and symmetric when compared with the thermodynamic maps of the cool gas galaxies (Figure \ref{fg:xrayent}). 
The  regular and thermally stable hot atmosphere extends to large radii, supporting the hypothesis that the energy injected by the AGN has cleared most of the cool gas from the core and has propagated further out, with the result of 
increasing the hot gas entropy.
NGC~4552, the other galaxy with no or very little cool gas reservoir, displays a hot atmosphere with similar thermodynamic properties but the X-ray gas morphology 
shows signs of imprints left by recent AGN energy release  \citep{machacek06b}.

Overall, our elliptical galaxies devoid of cool gas support the hypothesis that we are witnessing them during the more quiescent, overheated phase of the AGN feeding and feedback cycling, with both highly suppressed CCA rain and top-down condensation.\\

\noindent
{\bf Galaxies with a cool gas reservoir}
-- The spatial distribution and kinematics of CO and [CII] emitting gas
in NGC~4203, NGC~4261, and NGC~4636 are both in excellent agreement. 
The velocity structure measured in NGC~4203 and NGC~4261 indicates that the cold gas in these systems is distributed in rotating structures with well matched velocity and velocity dispersion in the two gas phases. Central dust absorption as revealed by {\it HST} data exhibits a similar spatial distribution.
Except for the lack of detected [CII] emitting gas, NGC~4374 reveals cold gas presence via the CO \citep{boizelle17} and dust emission.
The warm gas probed by the H$\alpha$+[NII] line emission roughly correlates with the spatial distribution of the colder component. 

The diversity of the galaxies in terms of environment in which they reside in and their dynamical status may induce the scatter seen in their thermodynamic profiles (Figure \ref{fg:thermo}), while still being consistent with variations driven by the AGN self-regulation cycle. 
The presence of cool gas in the fast rotator NGC~4203 could be enhanced by rotation. 
Indeed, cool gas can form and slowly accumulate in the central region because of the reduced feeding rate onto the SMBH due to the centrifugal barrier \citep{brighenti96, gaspari15}. For example, \citet{juranova19,juranova20} find favorable conditions for the development of multiphase gas condensations (with $C$-ratio\,$\sim 1$) in a sample of six massive ETGs with rotating gas, even with relatively large $t_{\rm cool}/t_{\rm ff}$ ratios.

 \begin{figure}[!t]
  \centering
  \includegraphics[width=8.2cm]{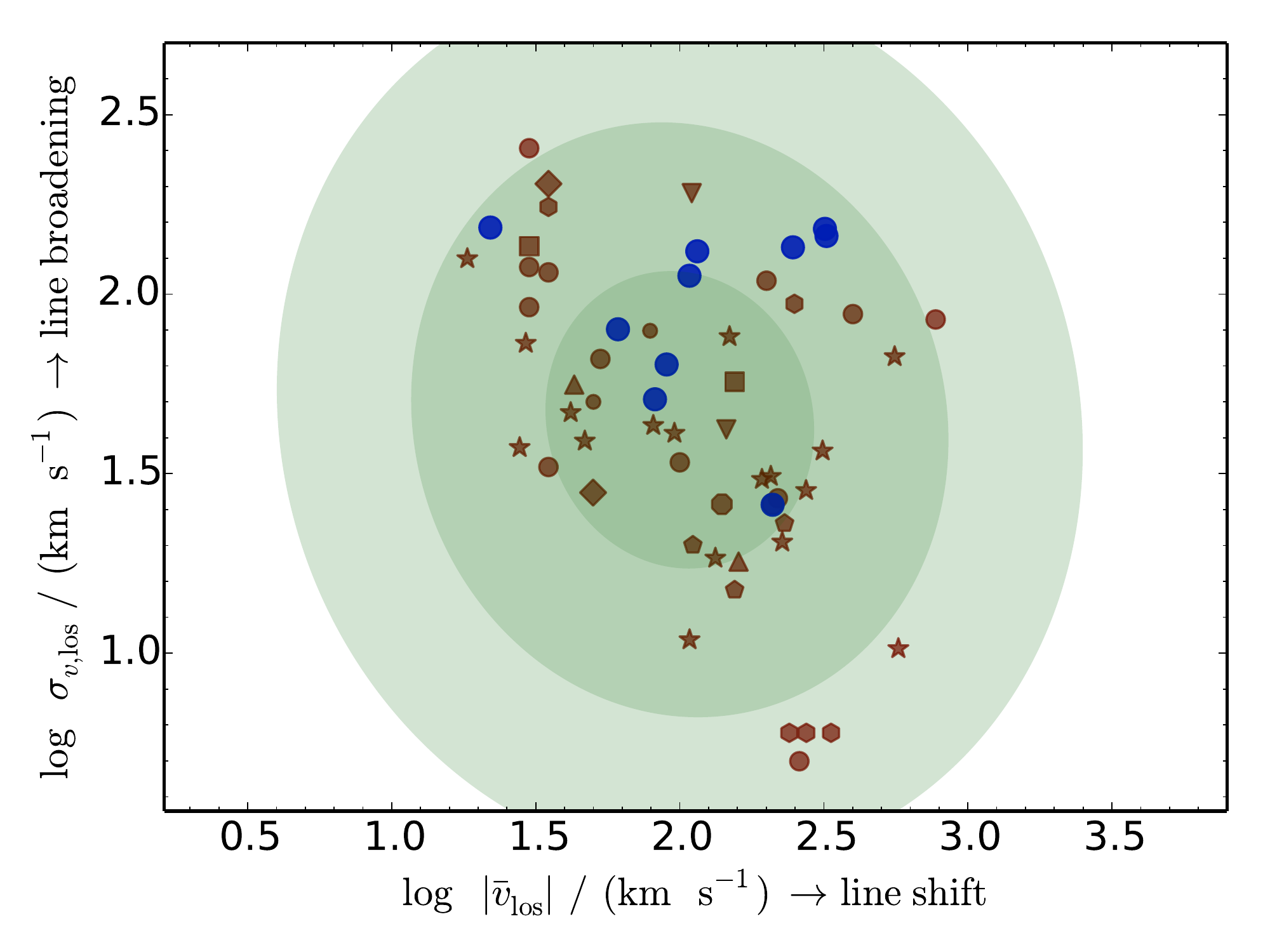}
  \caption{CCA k-plot diagnostic (pencil-beam): line broadening versus line shift, separating turbulence versus bulk motions. The red points are the observed cool clouds in massive ETGs from \citet{gaspari18} (see this paper for more details on the k-plot and their sample). The blue points denote the cool gas detected in our galaxies via [CII] and CO, whenever available (see Table \ref{tab:CII_CO_gal}; note that $\sigma_v = {\rm FWHM}/2.355$). For NGC 4261, we use only the more accurate two-Gaussian `2G' fit. The green shaded contours are the 1-3\,$\sigma$ confidence level predicted via high-resolution CCA simulations (\citealt{gaspari17_rain}). Our galaxies with cool gas are within the typically expected kinematics from a CCA-feeding cycle.}
  \vspace{+0.1cm}
  \label{fg:kplot}
\end{figure}

 \begin{figure}[!ht]
  \centering
  \includegraphics[width=8.cm]{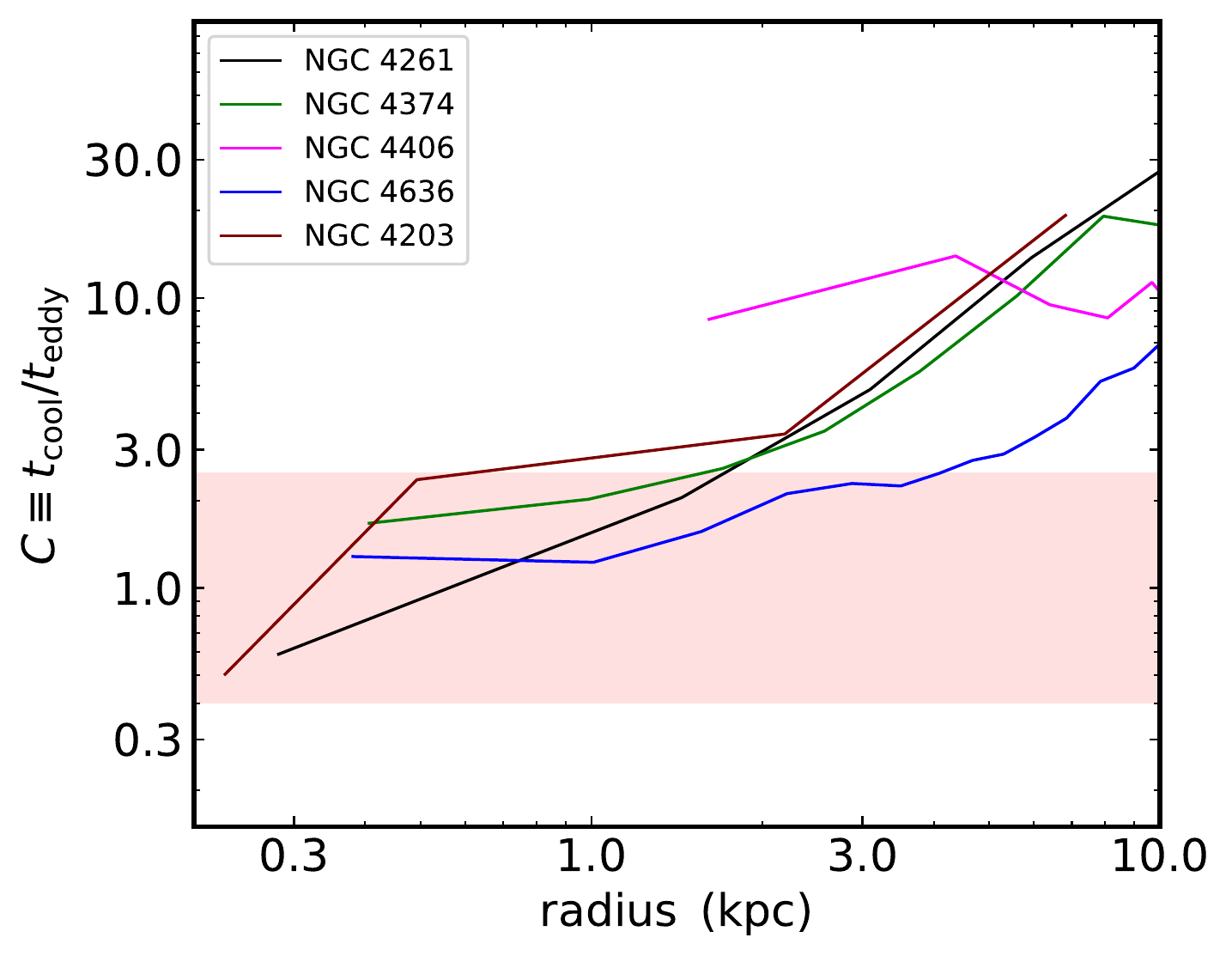}
  \caption{Ratio of the cooling time and turbulence eddy turnover time for our sample of galaxies with cool gas (see the legend and text for more details). The red band shows the 99\% confidence level predicted by the CCA simulations (\citealt{gaspari18}) within which significant nonlinear condensation is expected.
  Most of our non-central galaxies are expected to develop only an inner rain, at variance with the central NGC 4636 galaxy which can develop more extended multiphase condensation. Note also how the $C$-ratio is a more robust indicator of condensation, compared with $t_{\rm cool}/t_{\rm eddy}$, especially in the central regions. 
  }
  \vspace{+0.1cm}
  \label{fg:Cratio}
\end{figure}

Figure \ref{fg:kplot} shows one of the key CCA diagnostics, i.e., the line broadening versus the line shift in logarithmic space (\citealt{gaspari18}). This kinematical plot (`k-plot') is key to dissect the different cooling elements, in particular comparing the relative importance of turbulence (ordinate) versus bulk motions (abscissa). 
The blue points depict the detected cool gas in our galaxies via [CII] and CO, whenever available (see Table \ref{tab:CII_CO_gal}).
The green contours are the 1-3\,$\sigma$ confidence-level predictions from the simulated CCA condensed clouds in ETGs (\citealt{gaspari17_rain}). Such k-plot shows that the observed cool elements are within the expected kinematics from a CCA-feeding cycle, and are also comparable to those observed in several other galaxies (red points). 
A small deviation is that our sample shows slightly larger mean in the broadening axis, as we are including galaxies with rotation and AGN outbursts (e.g., NGC~4261), both shown to enhance the k-plot loci toward the upper-right sector (\citealt{maccagni21}). Moreover, our cool gas is often under-resolved (especially with SOFIA), meaning that we are only able to capture larger associations of clouds, rather than single elements, thus shifting the points toward the top region of the more ensemble-like regime, where clouds tend to drift in the macro-scale weather.

Another diagnostics of the top-down multiphase condensation is the $C$-ratio, i.e., the cooling time divided by the turbulence eddy time, $t_{\rm eddy} = 2\pi r^{2/3} L^{1/3}/\sigma_{v,L}$. Following \citet{gaspari18}, we compute the injection scale $L$ from either the diameter of the main AGN bubble, whenever available (NGC 4374, NGC 4636, NGC 4261; \citealt{finoguenov2008,baldi2009,o'sullivan11}), or the extension of the multiphase gas (roughly half the SOFIA beam). To get the macro-scale dispersion $\sigma_{v,L}$, we leverage the SOFIA FoV and average the related cool-gas velocity dispersions detected in each galaxy (Table \ref{tab:CII_CO_gal}; for NGC 4636 we also include \citet{werner14} multiphase gas observations). This is multiplied by $\sqrt{3}$ to account for the three-dimensional kinematics.
Figure \ref{fg:Cratio} shows the computed $C$-ratio for our galaxies with cool gas, superposed to the 99\% confidence-level band where significant precipitation is expected to trigger or be present.
NGC 4636 has the overall lowest $C$-ratio, which predicts a condensation rain up to $\sim$3 kpc, as corroborated by the observed extended multiphase filaments. 
Albeit the detection of the core is still missing, NGC 4406 has instead a $C$-ratio $\sim 10$, suggesting that the rain is very feeble or that the cool gas has originated from a merger (Sec.~\ref{s:external}).
In between, our three other galaxies have a $C$-ratio that is significantly below condensation threshold within $r<1$ kpc, thus expecting a more localized inner rain, consistently with the relative compact central structures observed in CO and [CII]. 
This shows how the condensation differs between central (NGC 4636) and non-central galaxies (the rest of our sample), the former trigger an extended rain, while the latter tend to develop a smaller sub-kpc weather region (as also predicted via X-ray scaling relations; cf.~figure 18 in \citealt{gaspari19}).

\begin{figure*}
  \centering
  \includegraphics[width=14.5cm]{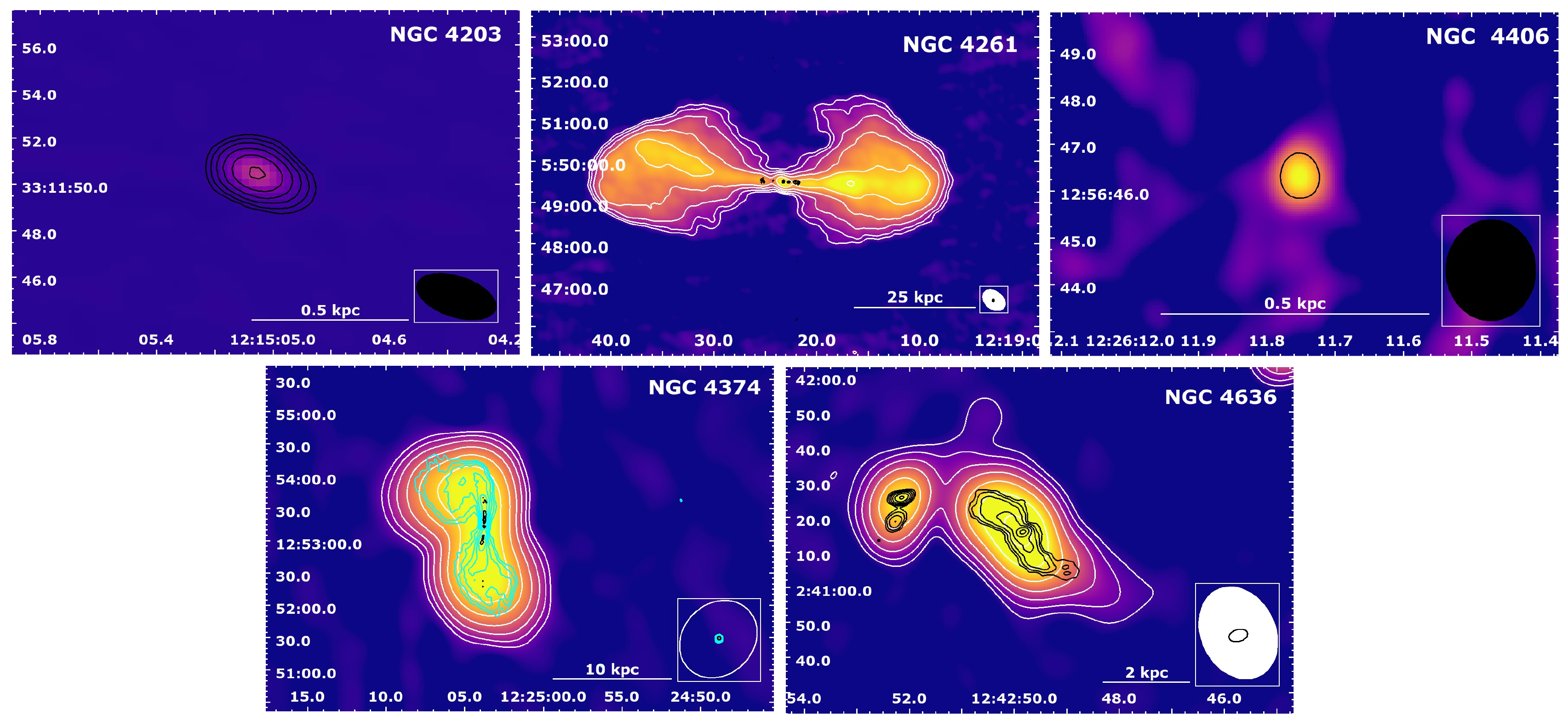}
  \caption{VLA radio total intensity maps at 1--2\,GHz for NGC\,4203 ({\bf top left}) and NGC\,4406 ({\bf top right}) in A configuration and NGC\,4261 ({\bf top middle}), NGC\,4374 ({\bf bottom left}) and NGC\,4636 ({\bf bottom right}) in the C configuration. The white radio contours corresponds to the VLA C configuration, the black contours to VLA A configuration and cyan contours to the VLA B configuration. In all cases, the contours are created at [1,2,4,8,16,...]$\times\,5$ RMS noise and the restoring beams (resolution) are shown in the bottom right corner of all radio maps (for more details see Table\,\ref{tab:radio_reduction_details}).}
  \label{fg:vla_gas}
\end{figure*}

Figure \ref{fg:vla_gas} shows the nuclear radio emission observed at 1--2\,GHz taken in several VLA configurations. The sources have a wide range of radio power and morphologies. 
Table \ref{tab:sum} presents the AGN jet power $P_{\rm jet}$,
calculated as the work required to inflate a cavity, taken from recent literature. 
The $P_{\rm jet}$ of the individual galaxies span a range of two orders of magnitude. 
The limited data available does not allow for a robust analysis of correlations between the $P_{\rm jet}$  and H$\alpha$ luminosity or the amount of cold gas. However, the three galaxies with (reliably) detected H$\alpha$+[NII] emission (NGC~4204, NGC~4374 and NGC~4636) seem to indicate a positive correlation of jet power with $L_{\rm H\alpha}$.
In the scenario of the recursive AGN self-regulation, such a correlation is expected since 
the SMBH accretion rate increases due to stronger CCA rain, giving rise to larger AGN power, while the cool gas mass increases via top-down condensation \citep{gaspari17_rain, gaspari18}.

The analysis of the radio power and the morphology of the radio emission in each individual galaxy may provide some insights into the AGN feeding/feedback processes.
Detailed studies of the very powerful jet in NGC~4261 \citep{o'sullivan11, kolokythas15} suggest that the radio source has gone through multiple AGN outbursts or significant changes in jet power with a time-scale of $\sim 10^5$ yr \citep{worrall10}. 
CCA-regulated feedback naturally predicts short AGN bursts for elliptical galaxies, which are proportional to the central cooling time \citep{gaspari19}. 
The related frequent chaotic AGN feeding and feedback cycles is likely responsible for 
the significant scatter in the cooling time and entropy profiles observed.

The large extent of the radio emission in NGC~4261 indicates that the energy deposition has progressively occurred at larger radii, allowing the cooling to occur at the center.
As a result the hot galactic
atmosphere is relatively undisturbed (Figure \ref{fg:xrayflux}) with fairly regular entropy distribution (Figure \ref{fg:xrayent}).
Radio observations of NGC~4374 and NGC~4636 show less powerful jets with energy deposited in the central region (NGC~4374) or just outside the radiance core (NGC~4636). Consistently, the X-ray gas shows distinct signs of disturbances generated by central energy deposition via AGN outbursts. 
Together with the k-plot and $C$-ratio above, the detection of our galaxies with cool gas overall corroborates the phase of the recursive CCA feeding, in which the increased condensation (in particular within the kpc region) triggers more vigorous AGN heating and activity.

\subsubsection{Stellar mass loss}
It is expected that cold gas originating from the cooling of the hot halo is initially dust poor \citep[e.g.][]{tsai95,valentini15}.
Given the relative short time scales for 
the grain sputtering time, 
$t_{\rm sp} \approx 1.2 \times 10^7 
(a / 0.1\mu{\rm m}) (n_e/10^{-2})^{-1}$ yr, where {\it a} is the grain radius \citep{temi2007b}, 
dust is not expected to survive long when in contact with the hot halo.
Although dust can grow by accretion of gas-phase metals in the cold gas, it is not clear
if the time scale of such process is compatible with the short sputtering time
\citep[e.g.][]{valentini15}. 
Thus, it is possible that some of the observed cool gas may originate from stellar mass loss. 
\citet{mathews03} showed that dusty 
gas ejected from red giant stars within about 1 kpc of the 
galaxy centers can settle into the core while retaining 
a significant fraction of its original dust. 
Dust is protected from sputtering if it remains in 
warm $T \approx10^4$ K ionized gas for 
$\sim10^6$ yrs, which is comparable to the dynamical 
(free-fall) time at $\sim$\,1 kpc in ellipticals. 
Alternatively, 
if this dust immediately goes into the $10^7$ K hot gas, it can 
rapidly cool it to $\sim 10^4$ K in $\lta 10^6$ yrs by inelastic 
electron-dust collisions \citep{mathews03}. 

Previous studies based on the detection of molecular gas in ETGs using the ATLAS and MASSIVE galaxy surveys\citep{young11,davis19} show no correlation between the molecular gas mass and the stellar mass. The molecular gas content in our galaxy sample, as shown in Figure \ref{fg:H2} and Table \ref{tab:CII_CO_gal}, covers approximately 2 orders of magnitude, while the $L_K$ luminosity is within a factor of two. Thus, these additional data confirm the lack of dependence of the molecular gas content on the stellar mass, with the implication that stellar mass loss is not the primary source of molecular gas in these galaxies.

\section{Conclusions} \label{conclusions}
We have investigated the cold gas content of a sample of 6  massive elliptical galaxies taken from
the sample of 18 nearby ($d<100$~Mpc), massive, X-ray bright early-type galaxies studied by \citet{dunn10}. New observations of the 157  $\mu$m CII emission line taken with the FIFI-LS instrument on board of SOFIA, along with new ALMA CO observations and radio data from VLA, are presented. These data, complemented with a large multiwavelength dataset available in literature, are used to probe the complex ISM of the galaxy sample. The results are compared to previous studies of elliptical galaxies in a  parent sample of 6 galaxies \citep{werner14} extracted from the same original sample of \citet{dunn10} and to the investigations of ETGs by the MASSIVE and ATLAS surveys.\\ 
We find that the gas/dust evolution in our galaxies (most of which are non-central galaxies)
seems to be driven by a combination of mechanisms which can alternate during the recent history of the galaxy. Several galaxies are well described by multiphase condensation diagnostics, while a few others (e.g., NGC 4406) display potential merger-like features. The central galaxies studied by \cite{werner14}, tend instead to be predominantly shaped by internal process, given the rare merger-like interactions. 
Given the deeper potential wells and larger gas mass, central ellipticals tend to have much lower cooling times ($\propto K^{3/2}$) over extended radii (tens kpc). 
These central galaxies are akin to cool-core systems, which have substantial condensation and AGN activity but on more extended regions. In contrast, our sample of non-central galaxies does not have such an extended gaseous halo, and as a consequence of lower densities and higher cooling times at large radii, the condensation occurs only within a few kpc. 
Overall, 
our key conclusions are summarized as follows:

\begin{itemize}
\item 
The small galaxy sample shows significant diversity in cold gas content.
The [CII] line emission is detected in three out of six galaxies. Complementary multiwavelength data of two galaxies undetected in CII with SOFIA FIFI-LS, NGC4649 and NGC4552, confirm that these systems are devoid of cold gas or significant amount of dust.
The third galaxy with no detection in the carbon line, NGC 4374 (M84) displays instead filamentary H$\alpha$+[NII] emission, dust, and molecular gas. Most likely, deeper observations in the 157 $\mu$m line may reveal CII emitting gas.

\item
The distribution of the detected cool gas phase in our six (non-central) galaxies is often centrally concentrated and compact, at variance with the cold gas-rich galaxies studied in \citet{werner14}. Although not large in absolute terms when compared to MASSIVE and ATLAS surveys of ETGs, the cold gas content spans two orders of magnitude among the sample.

\item
Although some of the observed cool gas may originate from stellar mass loss, the kinematical and photometric misalignment between cold gas and stars suggest that
in these galaxies, stellar mass loss is  not the primary source of molecular gas.

\item
 Given the diversified environment in which some of the galaxies live in, it is conceivable that  NGC~4261, NGC~4406, and NGC~4552 may have acquired part of their gas externally via galaxy-galaxy interactions, although we find at best qualitative evidences. 

\item
Most galaxies in our sample have similar entropy and cooling time profiles, albeit with large scatter. This is at variance with the sample of central massive ETGs studied by \citet{werner14}, with a dichotomy in the X-ray properties between cold-gas rich and poor systems. Combined with the diversity in the cold gas content, this argues for a composite origin of their multiphase medium or for a different stage in the AGN feeding/feedback cycle.

\item
A few galaxies have active radio jets, which tend to deposit their energy in the large-scale hot atmospheres. Gas cooling is promoted at small radii, while at larger radii the atmospheres remain single phase (except for NGC 4636), consistently with the observed spatial distribution of the cold gas.

\item
Comparing with the CCA predictions, we find that our cool gas-free galaxies are likely in the overheated interval of the self-regulated cycle without major signatures of recent AGN activity. For the cool-gas galaxies, the k-plot and AGN power correlation corroborate instead the phase of recursive CCA feeding in which the increased condensation (further enhanced via rotation) is starting to trigger more vigorous AGN heating. The~$C$-ratio consistently shows that central (NGC 4636) and non-central galaxies are expected to generate an extended or inner rain, respectively.

\end{itemize}

\section{\bf \scriptsize Acknowledgements}
\noindent
This work is based in part on observations made with the NASA/DLR Stratospheric Observatory for Infrared Astronomy (SOFIA). SOFIA is jointly operated by the Universities Space Research Association, Inc. (USRA), under NASA contract NNA17BF53C, and the Deutsches SOFIA Institut (DSI) under DLR contract 50 OK 0901 to the University of Stuttgart. 
This paper makes use of the following ALMA data: ADS/JAO.ALMA\#2015.1.01107.S, ADS/JAO.ALMA\#2017.1.00301.S, ADS/JAO.ALMA\#2017.1.00830.S. The National Radio Astronomy Observatory is a facility of the National Science Foundation operated under cooperative agreement by Associated Universities, Inc.
HPC resources were provided by the NASA High-End Computing (HEC) Program (SMD-1153) through the NASA Advanced Supercomputing (NAS) Division at Ames Research Center.
MG acknowledges partial support by NASA Chandra GO9-20114X and {\it HST} GO-15890.020/023-A, and the \textit{BlackHoleWeather} program.
NW and RG are supported by the GACR grant 21-13491X. RG thanks Elisabetta Liuzzo for supervision and guidance with ALMA data reduction.
AS is supported by the Women In Science Excel (WISE) programme of the Netherlands Organisation for Scientific Research (NWO), and acknowledges the World Premier Research Center Initiative (WPI) and the Kavli IPMU for the continued hospitality. SRON Netherlands Institute for Space Research is supported financially by NWO.

\software{CASA (v4.7.2, McMullin et al. 2007), ZAP (Soto et al. 2016), CIGALEMC (Serra et al. 2011) MAGPHYS14 (Da Cunha et al. 2008), XSPEC (Arnaud 1996), SPEX (Kaastra et al. 1996), scikit-learn (Pedregosa et al. 2011), GALFIT (Peng et al. 2010), photutils (Bradley et al. 2019)}

\bibliography{paper}
\bibliographystyle{biblio}

\end{document}